\documentclass[twocolumn,superscriptaddress,showpacs,preprintnumbers,amsmath,amssymb, pre]{revtex4}
\usepackage{graphicx}
\usepackage{bm}

\begin{document}
\title{The response of jammed packings to thermal fluctuations}

\author{Qikai Wu}
\affiliation{Department of Mechanical Engineering and Materials Science, Yale University, New Haven, Connecticut, 06520, USA}
\author{Thibault Bertrand} 
\affiliation{Laboratoire Jean Perrin UMR 8237 CNRS/UPMC, Universit\'e Pierre et Marie Curie, 75255 Paris Cedex, France}
\author{Mark D. Shattuck}
\affiliation{Department of Physics and Benjamin Levich Institute, The City College of the City University of New York, New York, 10031, USA}
\affiliation{Department of Mechanical Engineering and Materials Science, Yale University, New Haven, Connecticut, 06520, USA}
\author{Corey S. O'Hern}
\affiliation{Department of Mechanical Engineering and Materials Science, Yale University, New Haven, Connecticut, 06520, USA}
\affiliation{Department of Physics, Yale University, New Haven, Connecticut, 06520, USA}
\affiliation{Department of Applied Physics, Yale University, New Haven, Connecticut, 06520, USA}

\date{\today}

\begin{abstract}
We focus on the response of mechanically stable (MS) packings of
frictionless, bidisperse disks to thermal fluctuations, with the aim
of quantifying how nonlinearities affect system properties at finite
temperature.  In contrast, numerous prior studies characterized the
structural and mechanical properties of MS packings of frictionless
spherical particles at {\it zero temperature}.  Packings of disks with
purely repulsive contact interactions possess two main types of
nonlinearities, one from the form of the interaction potential
(e.g. either linear or Hertzian spring interactions) and one from the
breaking (or forming) of interparticle contacts. To identify the
temperature regime at which the contact-breaking nonlinearities begin
to contribute, we first calculated the minimum temperatures $T_{cb}$
required to break a single contact in the MS packing for both single
and multiple eigenmode perturbations of the $T=0$ MS packing. We find
that the temperature required to break a single contact for equal
velocity-amplitude perturbations involving all eigenmodes approaches
the minimum value obtained for a perturbation in the direction
connecting disk pairs with the smallest overlap.  We then studied
deviations in the constant volume specific heat $C_V$ and deviations
of the average disk positions $\Delta r$ from their $T=0$ values in
the temperature regime $T_{cb} < T < T_{r}$, where $T_r$ is the
temperature beyond which the system samples the basin of a new MS
packing.  We find that the deviation in the specific heat per particle
$\Delta {\overline C}_V^0/{\overline C}_V^0$ relative to the zero
temperature value ${\overline C}_V^0$ can grow rapidly above $T_{cb}$,
however, the deviation $\Delta {\overline C}_V^0/{\overline C}_V^0$
decreases as $N^{-1}$ with increasing system size.  To characterize
the relative strength of contact-breaking versus form nonlinearities,
we measured the ratio of the average position deviations $\Delta
r^{ss}/\Delta r^{ds}$ for single- and double-sided linear and
nonlinear spring interactions.  We find that $\Delta r^{ss}/\Delta
r^{ds} > 100$ for linear spring interactions is independent of system
size.  This result emphasizes that contact-breaking nonlinearities are
dominant over form nonlinearities in the low temperature range $T_{cb}
< T < T_r$ for model jammed systems.
\end{abstract}

\maketitle

\section{Introduction}
\label{introduction}

\begin{figure*}
\centering\includegraphics[width=0.9\textwidth,height=0.675\textwidth]{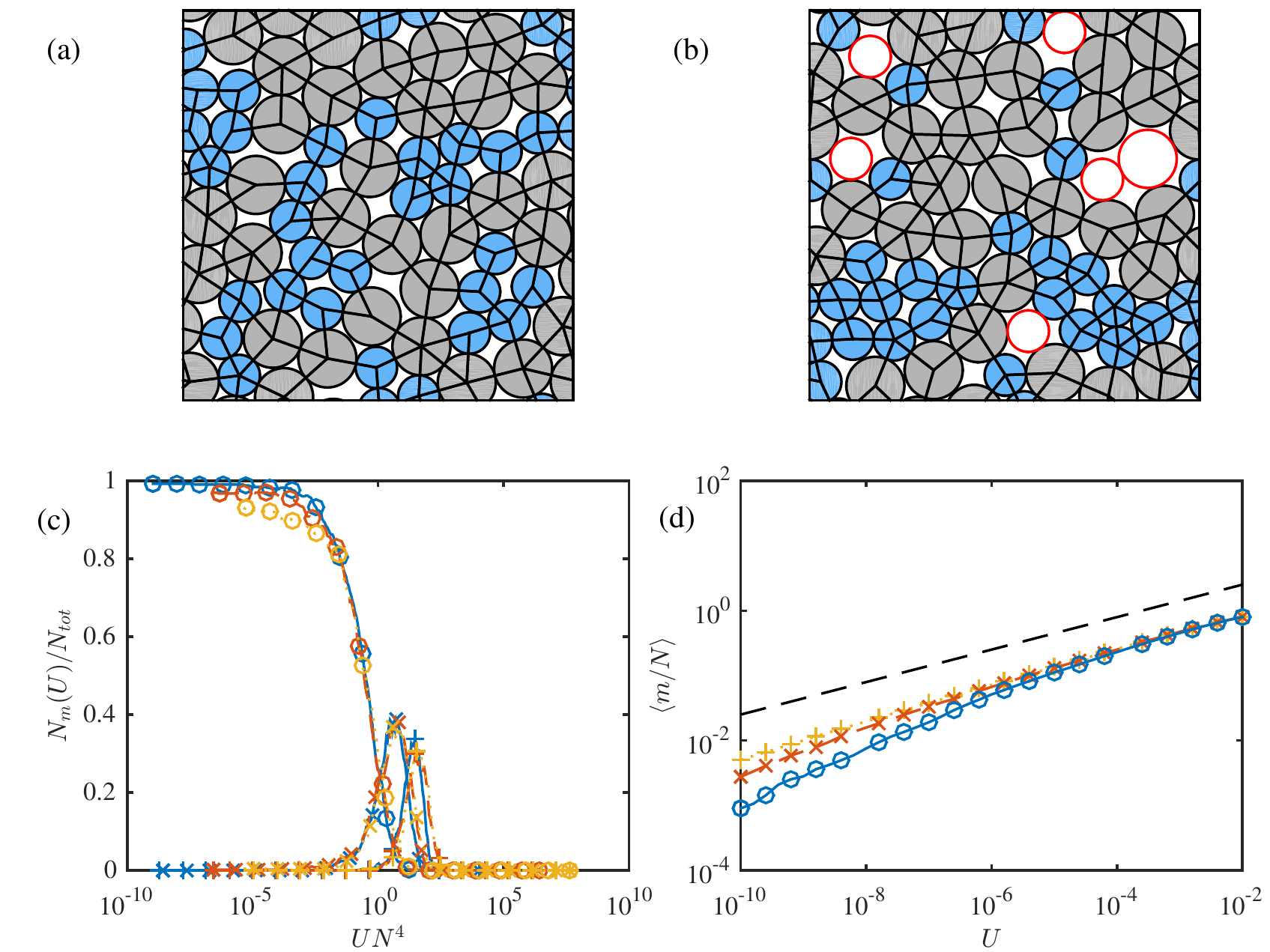}
\caption{Examples of isostatic mechanically stable bidisperse disk
packings at zero temperature with (a) $N_c = N_c^0 = 127$ contacts
and $N_r=0$ rattler particles and (b) $N_c=N_c^0=115$ contacts and
$N_r=6$ rattler particles. The non-rattler (rattler) disks are
outlined in black (red). For both (a) and (b), the total number of
disks $N=64$, the potential energy per particle $U = 10^{-12}$, and
the solid black lines connecting disks centers indicate
force-bearing interparticle contacts. (c) The fraction of mechanically stable
packings $N_m(U)/N_{\rm tot}$ that possess $m=N_c -N_c^0=0$
(circles), $2$ (exes), and $4$ (plus signs) excess contacts as a
function of $U N^4$ for three system sizes $N=32$ (solid lines), $128$
(dashed lines), and $256$ (dotted lines). (d) The number of
excess contacts $m$ normalized by $N$ averaged over $5000$
MS packings and plotted versus $U$ for three system sizes $N=32$
(circles), $128$ (exes), and $256$ (plus signs). The dashed line has
slope $0.25$.}
\label{fig1}
\end{figure*}

Static packings of frictionless disks and spheres are informative
model systems for studying jamming in granular media~\cite{behringer}
and dense colloidal suspensions~\cite{hunter}. Mechanically stable
(MS) packings of frictionless disks in two spatial dimensions (2D) are
isostatic at jamming onset~\cite{isostatic} and possess $N_c^0 = 2N' -
1$ contacts (with periodic boundary conditions), where $N'=N-N_r$ is
the number of disks in the force-bearing contact network, $N$ is the
total number of disks, and $N_r$ is the number of ``rattler" disks
with fewer than three contacts per disk~\cite{xu}. (See
Fig.~\ref{fig1} (a) and (b).) Mechanically stable disk packings
possess a full spectrum of $2N'-2$ nonzero eigenvalues of the
dynamical matrix (i.e. the Hessian of the interaction
potential~\cite{alexander1998amorphous}), which represent the
vibrational frequencies of the zero-temperature packings in the
harmonic approximation. The structural and mechanical properties of
isostatic disk and sphere packings near jamming at {\it zero
  temperature} have been reviewed
extensively~\cite{ohern,jam_review,jam_review2}, including the
pressure scaling of the bulk and shear moduli, excess contact number,
and low-frequency plateau in the density of vibrational modes near
jamming onset. More recently, several groups have investigated how the
scaling behavior of these quantities is affected by thermal
fluctuations using computer
simulations~\cite{ikeda2,ikeda2013dynamic,bertrand2,goodrich2}, and
mechanical vibrations in experiments of granular
media~\cite{dauchot,dauchot2,dauchot3}.

In particular, several authors have used computer simulations of soft
disks that interact via purely repulsive linear spring potentials to
study how the density of vibrational modes of mechanically stable
packings at zero temperature and finite overcompression (with
potential energy per particle $U>0$) changes with increasing
temperature.  This work has shown that there is a characteristic
temperature $T^* \sim U \sim \Delta \phi^2$, where $\Delta \phi =
\phi-\phi_J$ is the deviation in the packing fraction above jamming
onset at $\phi_J$, above which the density of vibrational modes begins
to deviate strongly from that at zero
temperature~\cite{ikeda2013dynamic,bertrand2,wang}.  In addition, they showed
that $T^*$ corresponds to the temperature above which an extensive
number of the contacts in the $T=0$ contact network has broken.
 
These prior studies emphasized that an extensive number of broken
contacts (or more~\cite{goodrich2}) were required to significantly
change the binned density of vibrational modes. However, do any
important physical quantities change when a single contact or
sub-extensive number of contacts in the zero-temperature contact
network is broken by thermal fluctuations?  The answer to this
question may depend on the number of excess contacts in the $T=0$
contact network $m = N_c - N_c^0$. For example, if a zero-temperature
packing has zero excess contacts ($m=0$), the breaking of a single
contact would cause the system to become unjammed. In Fig.~\ref{fig1}
(c), we show the fraction of MS packings with $m$ excess contacts,
$N_m(U)/N_{tot}$, can be collapsed for each $m$ and different system
sizes by plotting $N_m(U)/N_{tot}$ as a function of $U N^4$. We find
that the average number of excess contacts scales as $\langle m
\rangle/N \sim U^{1/4}$ (Fig.~\ref{fig1} (d)), which is consistent
with previous studies at zero temperature~\cite{ohern}. Thus, in the
large-system limit isostatic packings with $m=0$ exist only at $U=0$.

In this article, we will first characterize the minimum temperature
required to break a single contact as a function of the protocol used
to add thermal fluctuations.  We focus on this quantity because it can
be determined exactly in the low-temperature limit from the
eigenvalues and eigenmodes of the dynamical matrix for the $T=0$ MS
packings.  In particular, we will measure the minimum temperature
$T_1(m,m-1)$ above which a $T=0$ MS packing with $m$ excess contacts
changes to a packing with $m-1$ excess contacts in response to a
perturbation along a single eigenmode. Thermal fluctuations can also
be added to the zero-temperature MS packing by perturbing the system
along a superposition of $n$ eigenmodes of the dynamical matrix, and
we can measure the minimum temperature, $T_n(m,m-1)$, required to
break a single contact.  We will show that that the minimum
temperature required to break a single contact over all single mode
excitations scales as $T_1(m,m-1) \sim U/N^{\alpha}$, where $\alpha
\approx 2.6 \pm 0.1$, which is consistent with previous
measurements~\cite{schreck2}. For multi-mode excitations, $T_n(m,m-1)$
decreases as the number of eigenmodes $n$ involved in the perturbation
increases, reaching a minimum for perturbations with equipartition of
all $2N'$ eigenmodes.  The minimum temperature required to break a
single contact for a perturbation with equipartition of all eigenmodes
of the $T=0$ dynamical matrix scales as $T_{2N'}(m,m-1) \sim
N^{-\beta}$, where $\beta \approx 2.9 \pm 0.1$. This system-size size
dependence is stronger than that for single-mode perturbations.

We also measured the temperature required to break multiple contacts.
In this case, we employed molecular dynamics simulations to determine
the temperature at which a given fraction of simulation snapshots possess a
specified number of contacts.  This information cannot be
obtained from the $T=0$ dynamical matrix, since the eigenmodes
change after contacts begin breaking. We find a power-law scaling
relation between the temperature, number of broken contacts $N_{bc} =
N_c^0 + m - N_c$, system size, and potential energy per particle $U$.

After investigating the temperatures at which a given number of
zero-temperature contacts break, we search for physical quantities
that may be sensitive to changes in the interparticle contact
networks. We focused on two quantities: 1) the deviation in the
specific heat $\Delta C_V/C_V^0 = (C_V-C_V^0)/C_V^0$ from the zero
temperature value $C_V^0$ and 2) the deviation of the average
positions of the disks ${\overline x}_i$ and ${\overline y}_i$ in a
packing at a given temperature, $\Delta r = \sqrt{\sum_{i=1}^{N}
  [(\bar{x}_i - x_i^0)^2 + (\bar{y}_i-y_i^0)^2]/N}$, from the $T=0$
disk positions ${\vec R}^0 = \{
x_1^0,y_1^0,\ldots,x_{N'}^0,y_{N'}^0\}$. Calculating $\Delta r$ is
important for understanding how far the initial packing can move in
configuration space before transitioning to the basin of a new MS
packing. We compare $\Delta r^{ss}$ for systems with purely repulsive
(single-sided) linear and nonlinear spring interactions to $\Delta
r^{ds}$ obtained for systems with double-sided linear and nonlinear
spring interactions, which allows us to quantify the additional
nonlinearities that arise from contact breaking. We find that both
quantities, $\Delta C_V/C_V^0$ and $\Delta r^{ss}/\Delta r^{ds}$ are
sensitive to the breaking of a {\it single contact}. However, the
deviation $\Delta C_V/C_V^0$ decreases with increasing system size.
In contrast, $\Delta r^{ss}/\Delta r^{ds} > 100$ for purely repulsive
linear springs and does not depend strongly on system size.  We also
quantify $\Delta r^{ss}/\Delta r^{ds}$ for packings with Hertzian
spring interactions and show that contact breaking increases the
magnitude of the nonlinearities at finite temperature, but not as much
as for linear repulsive spring interactions.

\begin{figure}[h!]
\includegraphics[width=3.5in]{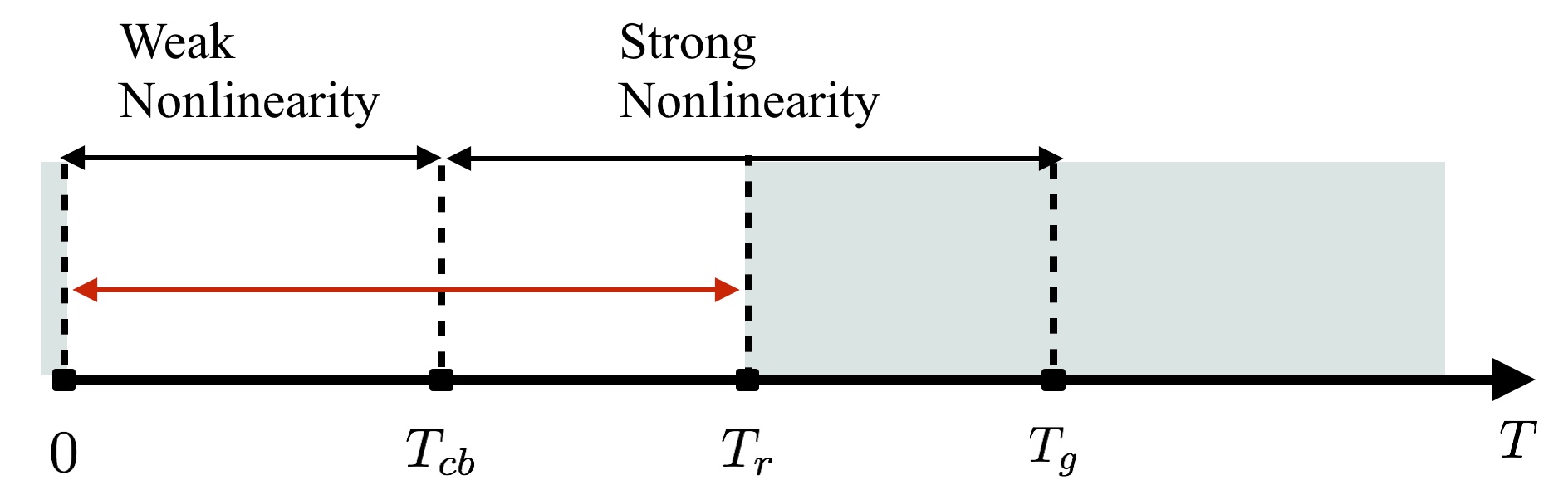}
\caption{Schematic of four important temperature regimes when studying
the response of MS packings to thermal fluctuations. For $T <
T_{cb}$, the $T=0$ contact network remains intact. In this regime,  
``form'' nonlinearities occur when the interparticle potential 
cannot be written exactly as a harmonic function of the disk 
positions. For $T_{cb} < T < T_r$,
the $T=0$ contact network changes, both form and contact-breaking 
nonlinearities occur, and the system remains in the
basin of attraction of the original MS packing.  For $T_r < T <
T_g$, the system can move to the basins of attraction of other MS
packings, but the relaxation times are sufficiently long that
structural relaxation is not complete.  For $T > T_g$, the system is
liquid-like with finite structural relaxation times.  This article
focuses on the (unshaded) temperature regimes that occur for $0 < T
\le T_r$. }
\label{schematic}
\end{figure}

There are several important temperature scales to consider when
studying the response of MS packings to thermal fluctuations.  In
Fig.~\ref{schematic}, we show four temperature regimes: $0 < T <
T_{cb}$, $T_{cb} < T < T_r$, $T_r < T < T_g$, and $T > T_g$. For $0 <
T < T_{cb}$, where $T_{cb}$ is the minimum temperature at which a
single contact breaks, the system is weakly nonlinear with ``form''
nonlinearities that arise when the interaction potential cannot be
expressed exactly as a harmonic function of the disk positions.  (We
use the notation $T_{cb}$ for the temperature required to break 
a single contact when we do not specify the type
of initial perturbation.) In the temperature regime $T_{cb} < T <
T_r$, contacts begin breaking, both form and contact-breaking
nonlinearities occur, and the system remains in the basin of
attraction of the original MS packing. In this regime $\Delta r^{ss}$
can be much larger than $\Delta r^{ds}$ due to contact-breaking
nonlinearities. At larger temperatures, $T_r < T < T_g$, the system
rearranges and moves beyond the basin of attraction of the original
$T=0$ MS packing, but the timescales are prohibitively long to allow
complete structural relaxation.  Finally for $T > T_g$, the system is
liquid-like with a finite structural relaxation time.

We emphasize that a number of studies have characterized the
structural and mechanical properties of MS packings at
$T=0$~\cite{torquato,goodrich,teitel}.  Further, many studies have
tracked the growth of the dynamical heterogeneities and the structural
relaxation times as $T \rightarrow T_g$ from above~\cite{shen,berthier,ozawa}.
However, few studies have focused on the low-temperature regimes $0 <
T < T_{cb}$ and $T_{cb} < T < T_r$, where the contribution of contact
breaking to the magnitude of the nonlinearities can be quantified at
finite temperature.  In future work, we will focus on the temperature
regime $T_r < T < T_g$ to understand the connection between the
geometry of the high-dimensional energy landscape and slow structural
relaxation.
 
The remainder of this article will be organized as follows. In
Sec.~\ref{methods}, we describe the methods we employ to generate
zero-temperature MS packings, the protocols used to add thermal
fluctuations to the MS packings, and the measurements of the changes
in the specific heat $\Delta C_V^0/C_V^0$ and average particle
positions $\Delta r$ of the packings from their $T=0$ values as a
function of temperature.  In Sec.~\ref{results}, we present our
results for $\Delta C_V/C_V^0$ and $\Delta r$. We show that $\Delta
C_V/C_V^0$ increases more strongly when a single contact in the $T=0$
MS packing changes.  We find that $\Delta C_V/C_V^0$ decreases with
increasing system size, however, the quantity $\Delta r^{ss}/\Delta
r^{ds}$, which identifies the distinct contribution of contact
breaking to the nonlinear response, does not depend strongly on system
size.  In Sec.~\ref{conclusion}, we summarize our results and
highlight promising future research directions that stem from this
work.  We also provide several Appendices that include additional
details of the methods and calculations we implement. In
Appendix~\ref{Tstar}, we provide additional details concerning the
method we used to calculate the minimum temperature required to break
a single contact with perturbations that involve $n$ eigenmodes of the
$T=0$ dynamical matrix with equal velocity amplitudes. In
Appendix~\ref{dr_flt}, we discuss the additional nonlinearities that
arise from rattlers in MS packings and affect $\Delta r(T)$ at finite
temperatures.  In Appendix~\ref{rearrange}, we describe the methods
that we employed to measure the rearrangement $T_r$ and glass
transition $T_g$ temperatures.  Finally, in Appendix~\ref{dr_1d}, we
show that the leading order term in the change in the average position
scales linearly with temperature, $\Delta r \sim T$, for a particle in
a one-dimensional cubic potential well.

\section{Methods}
\label{methods}

Our computational studies focus on measuring the response of MS
packings composed of $N$ bidisperse frictionless disks ($N/2$ large and $N/2$
small disks with diameter ratio $\sigma_L/\sigma_S=1.4$) to thermal
fluctuations with system sizes in the range from $N=16$ to $1024$ disks using
periodic boundaries in square simulation cells. The disks (all with
mass $m$) interact via the pairwise, purely repulsive potential,
\begin{equation}
\label{Ep}
U(r_{ij})=\frac{\epsilon}{\alpha} \left(1-\frac{r_{ij}}{\sigma_{ij}}\right)^{\alpha}\Theta\left(1-\frac{r_{ij}}{\sigma_{ij}}\right),
\end{equation}
where $r_{ij}$ is the separation between the centers of disks 
$i$ and $j$, $\sigma_{ij}=
(\sigma_i + \sigma_j )/2$ is the average disk diameter, $\epsilon$ is
the energy scale of the repulsive interaction, $\Theta(x)$ is the
Heaviside step function, and $\alpha = 2$ ($5/2$) corresponds to linear 
(Hertzian) repulsive spring interactions. We also consider disk packings  
that interact via double-sided 
spring potentials with a similar form to that in Eq.~\ref{Ep}:
\begin{equation}
\label{double}
U_{ds}(r_{ij})=\frac{\epsilon}{\alpha} \left|1-\frac{r_{ij}}{\sigma_{ij}}\right|^{\alpha}.
\end{equation}
For studies involving interactions in Eq.~\ref{double}, the
interparticle contact network is fixed to that in the $T=0$ MS packing
for all temperatures~\cite{schreck}.  Comparison of the results from
single- versus double-sided interactions allows us to determine the
strength of the nonlinearities that arise from contact breaking alone.

We generate MS packings as function of the total potential energy per
particle $U=\Sigma_{i>j} U(r_{ij})/N$ using a protocol that
successively compresses or decompresses the system in small packing 
fraction steps $\Delta \phi$ followed by conjugate gradient energy
minimization~\cite{xu}. The compression/decompression protocol is terminated
when the total potential energy per particle satisfies $|U_c-U|/U
< 10^{-16}$, where $U_c$ is the current and $U$ is the target 
potential energy per particle.

The initial perturbations will be applied along one or more of the
eigenmodes of the dynamical matrix of the $T=0$ MS packings. We denote
the $2N'-2$ non-zero eigenfrequencies of the dynamical matrix as
$\{\omega^1,\ldots,\omega^{2N'-2}\}$. Each eigenfrequency $\omega^i$
has an associated eigenvector $\hat{E}^i = \{e^i_{x1}, e^i_{y1},
e^i_{x2}, e^i_{y2},\ldots,e^i_{xN'},e^i_{yN'}\}$ that satisfies
$({\hat E}^i)^2=1$.  The disk velocities ${\vec V}^0=\{v^0_{x1},
v^0_{y1},\ldots, v^0_{xN'},v^0_{xN'}\}$ corresponding to the initial
perturbation can be expressed as a linear combination of the
eigenmodes of the dynamical matrix:
\begin{equation}
\label{v0}
{\vec V}^0 = \sum_{i=1}^{2N'-2}A_i\omega^i{\hat E}^i.
\end{equation}
We will use the notation that upper case vectors, e.g. ${\vec R}$ 
and ${\vec V}$, include both the particle and spatial dimensions, while 
lower case vectors, e.g. ${\vec r}$ and ${\vec v}$, only include the 
spatial dimensions. 

For sufficiently small amplitude perturbations, the time evolution of
the multi-particle velocities and positions are given in the harmonic 
approximation by
\begin{equation}
\label{vt}
\vec{V}(t) = \sum_{i=1}^{2N'-2}A_i\omega^i{\hat E}^i \cos(\omega^i t),
\end{equation}
and 
\begin{equation}
\label{rt}
\vec{R}(t) = {\vec R}^0 + \sum_{i=1}^{2N'-2}A_i{\hat E}^i \sin(\omega^i t),
\end{equation} 
where ${\vec R}^0$ gives the disk positions in the $T=0$ MS packing.
We calculate the temperature of the system using the average kinetic
energy per particle $K/N$~\cite{warr1995energy}.

For sufficiently large temperatures, when multiple $T=0$ contacts
break and new contacts form, we cannot use the $T=0$ eigenmodes of the
dynamical matrix to determine the properties of the contact
networks. Thus, we will characterize the relation between the
temperature, number of contacts, system size, and potential energy per
particle using molecular dynamics simulations at constant number of
disks, area, and total energy $E$.  For the MD simulations, we use the
velocity Verlet integration scheme with a time step $\Delta t \sim
t_{\rm col}/40$, where $t_{\rm col} = \sigma_S \sqrt{\epsilon/m}$ is a
typical interparticle collision time scale, which provides total
energy conservation with relative standard deviation $\delta E/E <
10^{-13}$.

To investigate the effects of contact breaking, we will measure two
physical quantities as a function of the amplitude (or temperature) of
the thermal fluctuations.  We will first
study the change in the constant volume specific heat $\Delta C_V$
from its zero-temperature value $C_V^0 = 2 N' k_b$:
\begin{equation}
\label{dCv}
 \frac{\Delta C_V(T)}{C_V^0}= \frac{C_V(T)-C_V^0}{C_V^0},
\end{equation}
where $C_V = dE/dT$ and $k_b$ is the Boltzmann constant. In the 
low-temperature limit,
Eqs.~\ref{vt} and~\ref{rt} can be used to calculate the total energy:
\begin{equation}
\label{E}
E=K(t)+{\cal U}(t)={\cal U}_0+\frac{m}{2}\sum_{i=1}^{2N}A_i^2\omega_i^2={\cal U}_0+2N'k_bT,
\end{equation}
where ${\cal U}_0$ is the initial total potential energy. We will
measure the specific heat per particle ${\overline C}_V$ in the molecular
dynamics simulations by taking the temperature derivative numerically,
\begin{eqnarray}
\label{Cv} 
& & {\overline C}_V(T) = \\
&& \frac{1}{N'} \frac{E(T+dT)-E(T)}{dT} =
k_b \frac{E(T+dT)-E(T)}{K(T+dT)-K(T)}. \nonumber
\end{eqnarray}
From Eq.~\ref{Cv}, the deviation in the specific heat per particle can 
be written as  
\begin{equation}
\label{dCv2}
\Delta {\overline C}_V(T)/{\overline C}_V^0 =\frac{1}{2} \frac{E(T+dT)-E(T)}{K(T+dT)-K(T)}-1.
\end{equation}

We will also quantify the changes in the average positions of the
disks, $\Delta r$, as a function of temperature. We define $\Delta r$
as
\begin{equation}
\label{dr}
 \Delta r(T) = \sqrt{\frac{1}{N'} \sum_{i=1}^{N'} \left[(\bar{x}_i(T) - x_i^0)^2 + (\bar{y}_i(T)-y_i^0)^2\right]},
\end{equation}
where $(x_i^0, y_i^0)$ are the $x$- and $y$-coordinates of the $i$th
disk in the $T=0$ MS packing and $(\bar{x}_i(T), \bar{y}_i(T))$ are
the time-averaged $x$- and $y$-coordinates of disk $i$ at temperature
$T$. $\Delta r(T)$ can be interpreted as the average distance in the
$2N'$-dimensional configuration space between the $T=0$ MS packing and
the packing at finite $T$.  Note that rattler disks are not included  
in the calculations of $\Delta r$. 

\begin{figure}[h!]
\includegraphics[width=3in]{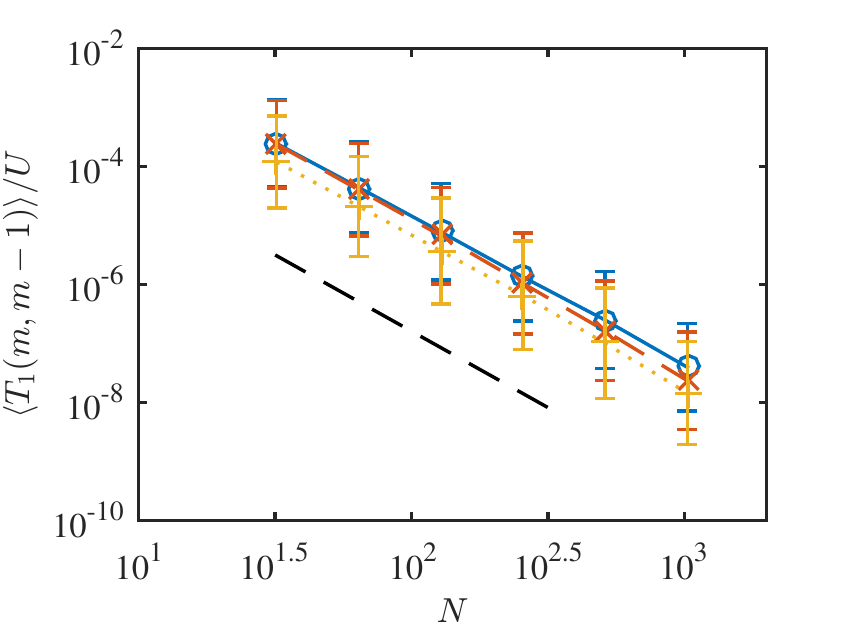}
\caption{The minimum temperature $T_1(m,m-1)$ required to break a
single contact when perturbing an MS packing along
one of the eigenmodes of the dynamical matrix averaged over $5000$
MS packings, normalized by the potential energy per particle $U$, and
plotted as a function of system size $N$.  $T_1(m,m-1)$ was obtained
by minimizing over all single mode perturbations.  We include
results for $U=10^{-12}$ (circles), $10^{-8}$ (exes), and $10^{-4}$
(pluses). The slope of the dashed line is $-2.6$. Rattler disks
are removed from the packings prior to performing these
calculations.}
\label{singlemode}
\end{figure}

\section{Results}
\label{results}

We organize our results into two main sections. In Sec.~\ref{Tcb}, we
discuss the results for the minimum temperatures required to break one
or more contacts for single- and multi-mode perturbations.  In
Sec.~\ref{quantities}, we show our results for the temperature dependence 
of the deviation in the specific heat per particle $\Delta {\overline C}_V$
and deviation in the average disk positions $\Delta r$ from those 
in the $T=0$ MS packing as a function 
of temperature.  For $T < T_{cb}$, form nonlinearities give rise 
to non-zero values of $\Delta {\overline C}_V$ and $\Delta r$. For 
$T > T_{cb}$, both form and contact-breaking nonlinearities are 
present. By comparing $\Delta {\overline C}_V$ and $\Delta r$ for 
single- and double-sided spring interactions, we can isolate the 
effects of the contact-breaking nonlinearities. We find that for $T_{cb} < T 
< T_{r}$
the specific heat deviation $\Delta {\overline C}_V$ scales as $N^{-1}$, 
whereas $\Delta r$ is roughly independent of system size.  
 
\begin{figure*}
\includegraphics[width=0.9\textwidth,height=0.675\textwidth]{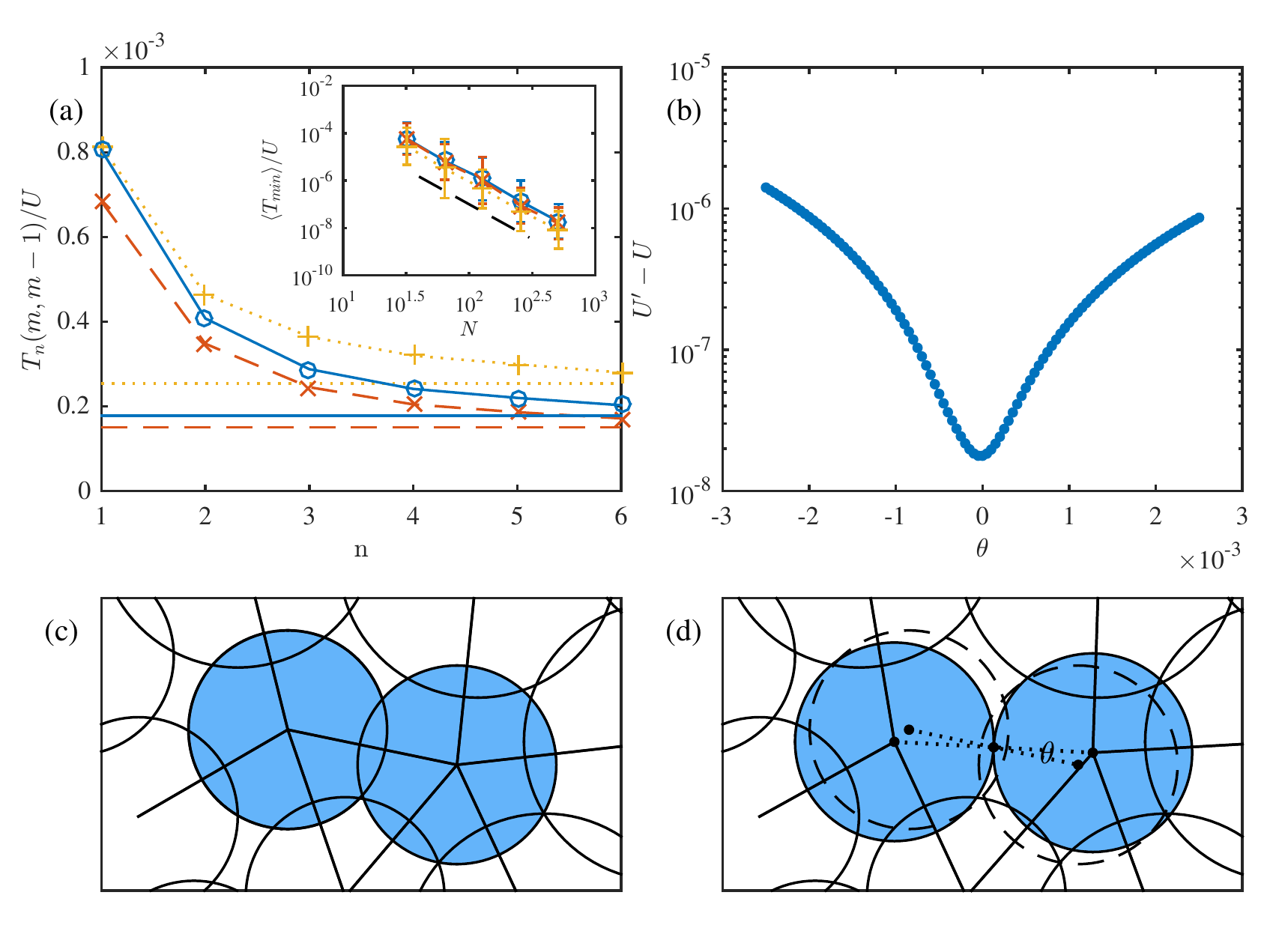}
\caption{(a) The minimum temperature $\langle T_n(m,m-1) \rangle$ 
(normalized by $U$ and averaged over $5000$ MS packings)
required to break a single contact in response to perturbations
that include $n=1,2,\ldots,6$ eigenmodes of the dynamical
matrix. $\langle T_n(m,m-1) \rangle$ is obtained by minimizing over all possible
$n$-mode combinations of the $2N'-2$ eigenmodes for each MS packing
at $U=10^{-12}$ (dashed line), $10^{-8}$ (solid line), and $10^{-4}$ (dotted line). 
The horizontal lines give the minimum temperature
$\langle T_{min}/U \rangle$ required to remove the smallest overlap between a pair
of contacting disks at each $U$ (averaged over $500$ MS packings). The inset shows the scaling of
$\langle T_{min}/U \rangle$ with system size $N$ for the same values of $U$ as 
in the main panel. The slope of the dashed
line is $-2.9$. (b) Difference in the potential energy per particle
between MS packings before ($U$) and after ($U'$) separating the pair 
of disks with the smallest
interparticle overlap as a function of the angle $\theta$ between
the old and new separation vectors between the two disks.  
(c) and (d) Schematic of the process to measure $T_{min}/U$.  In panel (c), the disk pairs with the 
smallest overlap are shaded in blue. In panel (d), this pair of 
disks is shifted so that $r_{ij}=\sigma_{ij}$.  The original positions are indicated by the 
dashed circles. The new separation vector makes an angle $\theta$ with 
the old separation vector (as indicated by the dotted lines). After shifting disks $i$ and $j$, potential 
energy minimization is performed allowing all disks to move except 
$i$ and $j$.
In both panels, the contact networks of the blue-shaded disks are 
indicated by solid lines.}
\label{mixmode}
\end{figure*}

\subsection{Temperatures required to break single and multiple contacts}
\label{Tcb}

In this section, we study the minimum temperature required to break a 
given number of contacts in the $T=0$ MS packing. We first focus on the 
breaking of a single contact and then study the breaking of multiple 
contacts.  We will show that the temperature required to
break the first contact depends strongly on the form of the initial
perturbation.  For example, the minimum temperature is smaller for 
perturbations along multiple eigenmodes compared to the minimum temperature 
for perturbations along a single eigenmode. 

At sufficiently low temperatures, we can use the harmonic
approximation for the disk positions given in Eq.~\ref{rt} to
calculate exactly the minimum temperature required to break a single
contact.  If we introduce a perturbation along a single eigenmode $k$,
the minimum temperature required to break a single contact
$T^k_1(m,m-1)$ can be calculated by first solving
$r_{ij}^2=\sigma_{ij}^2$ for all contacting disk pairs $i$ and $j$ and
then finding the minimum perturbation amplitude (or temperature) over
all disk pairs:
\begin{eqnarray}
\label{Tij}
& & T^k_1(m,m-1) = \\
& & \min_{i>j}\left\{\left[\frac{|\vec{\delta}_{ij}^k \cdot \vec{r}^0_{ij}|}{|\vec{\delta}^{ij}_k|^2} \left(\sqrt{1+\frac{(\sigma_{ij}^2-|\vec{r}^0_{ij}|^2)|\vec{\delta}^{ij}_k|^2}{|\vec{\delta}^{ij}_k \cdot \vec{r}^0_{ij}|^2}}-1\right)\right]^2\right\} \nonumber,
\end{eqnarray}
where $\vec{\delta}_{ij}^k=\vec{e}_{ij}^k \sin(\omega^k t)/\omega^k$
and $\vec{e}_{ij}^k = (e_{xi}^k-e_{xj}^k, e_{yi}^k-e_{yj}^k)$.  To
calculate the minimum $T_1^k(m,m-1)$ over all eigenmodes, we set
$|\sin(\omega^k t)|=1$ and find $T_1(m,m-1) = \min_k
T_1^k(m,m-1)$. (See additional details in Appendix~\ref{Tstar}.)

In Fig.~\ref{singlemode}, we show $\langle T_1(m,m-1) \rangle/U$
averaged over $5000$ MS packings as a function of system size $N$ for
three values of $U$. We find that $\langle T_1(m,m-1) \rangle$
normalized by $U$ collapses the data and $\langle T_1(m,m-1)\rangle/U$
displays power-law scaling with system-size, $\langle T_1(m,m-1)
\rangle/U \sim N^{-\alpha}$, where $\alpha \approx 2.6 \pm 0.1$. Thus,
$\langle T_1(m,m-1) \rangle$ tends to zero in the large system
limit~\cite{schreck2}, which stems from the increasing probability
for MS packings to possess anomalously small overlaps as $N
\rightarrow \infty$.

We now consider multi-mode perturbations and measure the 
minimum temperature required to break a single contact in 
$T=0$ MS packings. If we include $n$ eigenmodes in the perturbation, 
in the low-temperature limit, the disk positions and velocities are given by 
\begin{equation}
\label{vt_n}
\vec{V}(t) = \sum_{k=1}^{n}A_{k}\omega^{k}{\hat E}^{k}\cos(\omega^{k} t),
\end{equation}
and
\begin{equation}
\label{rt_n}
\vec{R}(t) = \vec{R}^0 + \sum_{k=1}^{n}A_{k}{\hat E}^{k}\sin(\omega^{k} t).
\end{equation} 

As for the single eigenmode perturbations, we can use the harmonic
expression for ${\vec R}(t)$ (Eq.~\ref{rt_n}) to determine the minimum
temperature required to break a single contact for multi-mode
perturbations.  Setting $r_{ij}^2=\sigma_{ij}^2$ for each pair of
disks in the force-bearing backbone yields an expression similar to
that in Eq.~\ref{Tij}, except $\vec{\delta}^{ij}_k$ is replaced by
$\vec{\delta}^{ij}=\sum_{k=1}^{n}\vec{e}_{ij}^{k} \sin(\omega^{k}
t)/\omega^{k}$.  The minimum temperature required to break a single
contact is obtained by evaluating the extrema of the sine functions,
$|\sin(\omega^{1} t)|=|\sin(\omega^2 t)|=\ldots=|\sin(\omega^{n}
t)|=1$, where we must check all combinations of $\sin(\omega^k t) =
\pm 1$, and by minimizing over all contacting disk pairs. For small
$n$, we discretize all of the possible eigenmode amplitude ratios
between $10^{-2}$ and $10^2$ and identify the amplitude ratio
combination that yields the minimum temperature $T_n(m,m-1)$ to break
a single contact.  For $n=2$ and $3$, we explicitly showed that
$T_n(m,m-1)$ is minimized (over all possible 
perturbations) for equal velocity-amplitude perturbations.
For $n >3$, we assumed that $A_1 \omega^1=A_2 \omega^2=\ldots=A_k
\omega^k$ perturbations give the minimum $T_n(m,m-1)$.  (Additional
details concerning these calculations are included in
Appendix~\ref{Tstar}.)

In Fig.~\ref{mixmode} (a), we plot $T_n(m,m-1)/U$ for single MS
packings using multi-mode perturbations as a function of the number of
eigenmodes $n=1,2,\ldots,6$ for three values of $U$, $10^{-12}$,
$10^{-8}$, and $10^{-4}$. For all $U$, we find that $T_n(m,m-1)/U$
decreases with increasing $n$ and then begins to saturate for $n
\gtrsim 6$.  In general, the minimum temperature required to break a
single contact decreases with an increasing number of eigenmodes in the
perturbation because the perturbation is more likely to have a
signficant projection onto the separation vector corresponding to the
smallest overlap between disks. Saturation of $T_n(m,m-1)$ with
increasing $n$ is interesting because it implies that the probability
to obtain a pair of disks in the force-bearing backbone with vanishing
overlap is zero in any finite-sized system with $U>0$.

We also developed a method to estimate $T_n(m,m-1)$ in the large-$n$
limit, which is illustrated in Fig.~\ref{mixmode} (c) and (d). We
first identify the pair of disks $i$ and $j$ in the force-bearing
backbone with the smallest overlap. We separate disks $i$ and $j$ so
that $r_{ij} = \sigma_{ij}$, while maintaining the center of mass of
the two disks and fixing all of the positions of the other disks in
the MS packing. We then minimize the total potential energy, allowing
all disks to move except disks $i$ and $j$, as a function of the angle
$\theta$ between the old and new separation vectors before
minimization.  In Fig.~\ref{mixmode} (b), we plot the difference
$U'-U$ in the potential energy per particle before ($U$) and after
($U'$) shifting disks $i$ and $j$ and minimizing the potential energy
as a function of $\theta$.  We find that the $\theta=0$ direction
gives rise to the smallest energy barrier, and thus we define the
temperature scale $T_{min} = U'(\theta=0) - U$.  $T_{min}/U$ provides
an accurate estimate of the large-$n$ plateau value of $T_n(m,m-1)/U$.
(See Fig.~\ref{mixmode} (a).) In the inset of Fig.~\ref{mixmode} (a),
we show $\langle T_{min}\rangle/U$ averaged over $500$ MS packings as
a function of system size.  $\langle T_{min} \rangle/U \sim \langle
T_n(m,m-1)\rangle/U \sim N^{-\beta}$, where $\beta \approx 2.95 \pm
0.05$, and displays stronger system-size dependence than $\langle
T_1(m,m-1)\rangle/U$.

\begin{figure}[h!]
\includegraphics[width=3in]{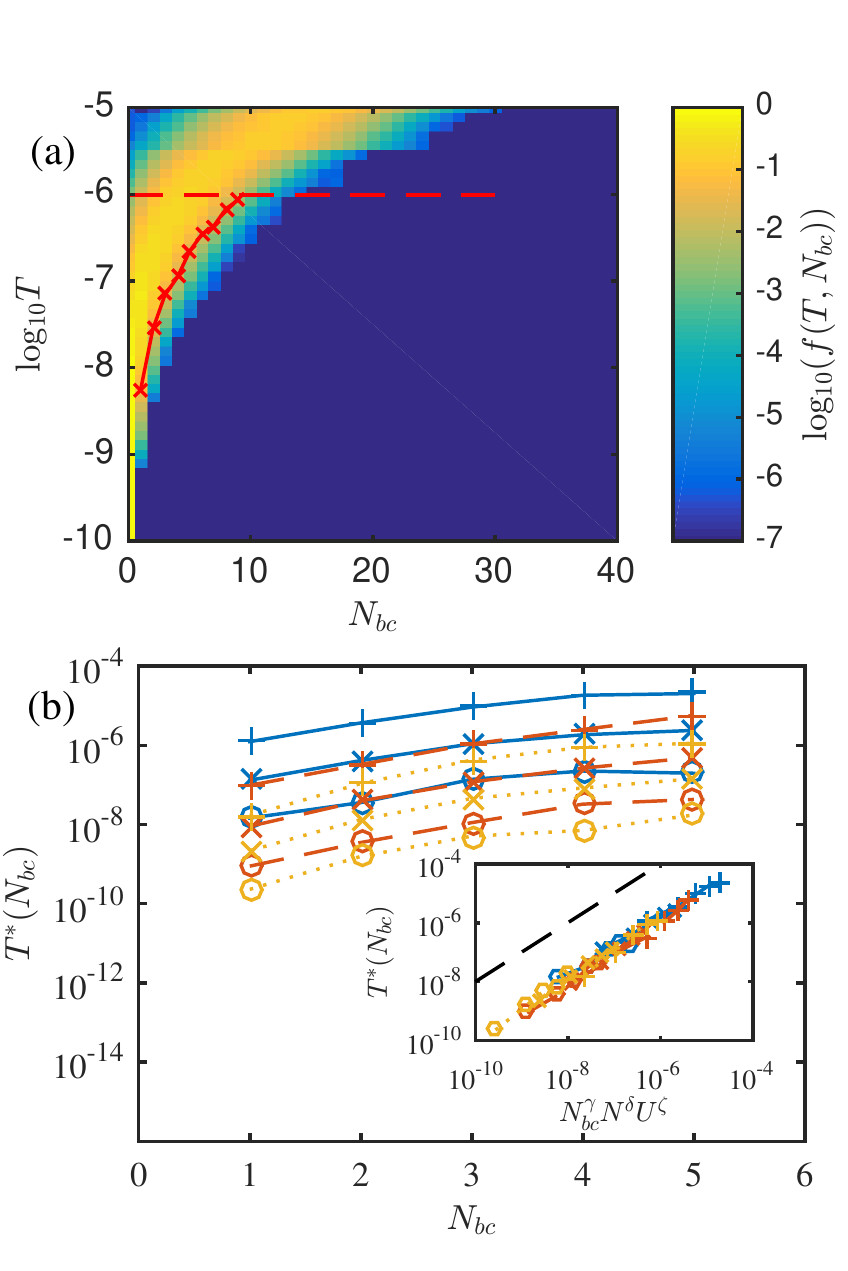}
\caption{(a) The fraction of time $f(T, N_{bc})$ that the system 
(with $N=64$ and $U=10^{-4}$) 
possesses a given number of broken contacts $N_{bc} = N_c^0 + m - N_c$ 
at temperature $T$. The color scale from yellow to blue represents 
decreasing $f$ on a $\log_{10}$ scale. The horizontal line indicates the 
rearrangement temperature $T_r$. The solid curve with exes gives the  
characteristic temperature $T^*(N_{bc}) < T_r$ for multiple contact 
breaking for which the fraction 
$f=0.1$. (b) The
characteristic temperature $T^*(N_{bc})$ for three system sizes,
$N=32$ (solid lines), $64$ (dashed lines), and $128$ (dotted lines), and 
three values of $U$, $10^{-5}$ (circles), $10^{-4}$ (exes), and 
$10^{-3}$ (pluses), for each $N$. The inset shows the same data as
in the main panel, but $T^*$ is plotted as a function of $N_{bc}^{\gamma}
N^{\delta} U^{\zeta}$,
where $\gamma \approx 2.2 \pm 0.3$, $\delta \approx -2.2 \pm 0.2$, and 
$\zeta \approx 1.0 \pm 0.1$.  The slope
of the dashed line is $1$.}
\label{multicontact}
\end{figure}

Thus far, we have focused on the minimum temperature $T_{cb}$ required
to break a single contact in $T=0$ MS packings for different forms of
the initial perturbations. For these calculations, we used the
harmonic approximation to determine the time-dependent disk positions
following the perturbation. We now consider temperatures beyond which
multiple $T=0$ contacts can break and new contacts can form.  As
discussed previously in Ref.~\cite{schreck2,bertrand2}, the eigenmodes
and associated eigenvectors can change significantly from those at
$T=0$ for $T > T_{cb}$, where new contacts can form and contacts at
$T=0$ can break.  Thus, for multiple contact breaking, we use constant
energy molecular dynamics simulations to directly measure the number
of contacts as a function of time following equal velocity-amplitude
perturbations. For these studies, we remove rattler disks prior 
to starting the simulations and focus on the temperature range $T < T_r$. 

During long trajectories, we measure the fraction of time
$f(T,N_{bc})$ at each temperature $T$ that the system possesses a
given number of broken contacts $N_{bc} = N_c^0 + m - N_c$. We show
$f(T,N_{bc})$ for a system with $N=64$ and $U=10^{-4}$ in
Fig.~\ref{multicontact} (a). At low temperatures $T < 10^{-9}$, $f$ is
large only for $N_{bc}=0$. As $T$ increases, more configurations
possess an increasing number of broken contacts. We can define a
characteristic temperature $T^*(N_{bc})$ for multiple contact breaking
by setting $f(T,N_{bc})=0.1$. 

In Fig.~\ref{multicontact} (b), we show the characteristic temperature
$T^*(N_{bc})$ for three system sizes $N$ and three values of the
initial potential energy per particle, $U$, for each $N$. We find that
$T^*$ obeys the following scaling form: $T^* \sim N_{bc}^{\gamma}
N^{\delta} U^{\zeta}$, where the exponents $\gamma \approx 2.2 \pm
0.3$, $\delta \approx -2.2 \pm 0.2$, and $\zeta \approx 1.0 \pm 0.1$.
(Other thresholds $0 < f <0.1$ give similar values for the exponents
$\gamma$, $\delta$, and $\zeta$.)  The scaling form suggests that
$T^*/U \sim N^2_{bc}/N^2 \sim (\Delta m/N)^2$, where $\Delta m$ is
difference in the excess number of contacts at $T=0$ and finite $T^*$.
This result shows that the temperature required to break an extensive
number of contacts scales quadratically with the change in the number
of contacts per particle in the range $0 < T < T_r$, which 
is consistent with prior results~\cite{ikeda2013dynamic,bertrand2,wang}.

\begin{figure}[h!]
\includegraphics[width=3in]{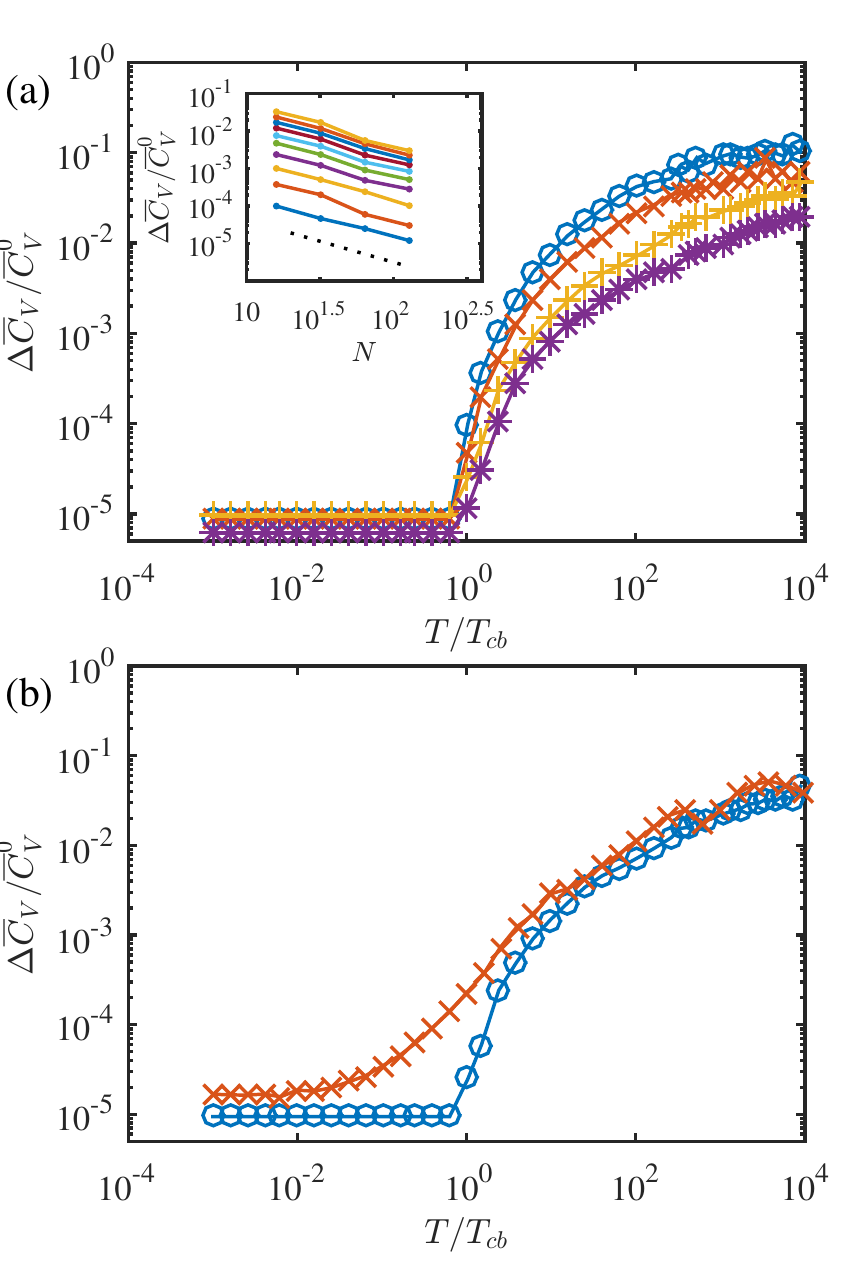}
\caption{(a) The normalized deviation in the specific heat per
particle at constant volume from the value at $T=0$, $\Delta
{\overline C}_V/{\overline C}_V^0$, at $U=10^{-5}$ for purely
repulsive linear spring interactions as a function of temperature
$T$ normalized by $T_{cb}$, where the first contact breaks. The data
is obtained from MD simulations at constant energy following equal
velocity-amplitude perturbations applied to $50$ $T=0$ MS packings
with $N=16$ (circles), $32$ (exes), $64$ (pluses), and $128$
(stars). The inset shows $\Delta {\overline C}_V/{\overline C}_V^0$
versus system size $N$ for $10$ values of $T/T_{cb}$ from $1$ to
$10^2$ (from bottom to top). The dotted line has slope $-1$. (b)
$\Delta {\overline C}_V/{\overline C}_V^0$ as a function of $T/T_{cb}$ for
MS packings with $N=32$ disks that interact via purely 
repulsive linear (circles)
and Hertzian spring interactions (exes) at $U=10^{-5}$.}
\label{dCv}
\end{figure}

\begin{figure*}
\includegraphics[width=0.9\textwidth,height=0.675\textwidth]{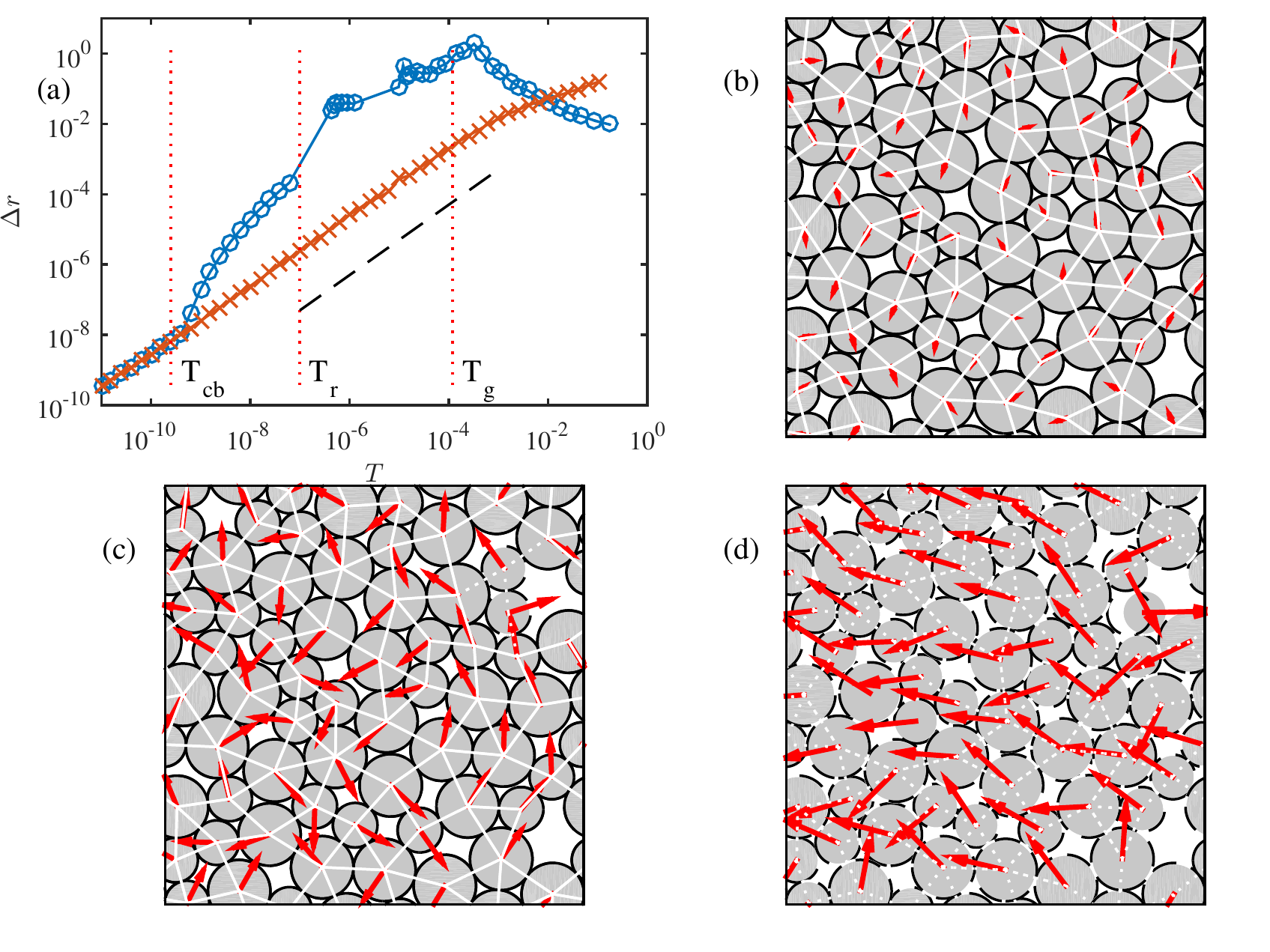}
\caption{(a) The deviation $\Delta r$ in the disk positions from
their $T=0$ values as a function of temperature $T$ for a MS packing
with $N=64$ and $U=10^{-5}$. The data is obtained from constant energy 
MD simulations with equal velocity-amplitude initial perturbations
involving all eigenmodes. We 
consider both purely repulsive
linear spring interactions (circles; $\alpha = 2$ in Eq.~\ref{Ep})
and double-sided linear spring interactions (exes; $\alpha =2$ in
Eq.~\ref{double}). The dashed line has slope $1$. The three dotted
vertical lines indicate 1) the measured temperature $T_{cb}$ at which the first
contact breaks, 2) the temperature $T_r$ at which the system
transitions to the basin of a new MS packing, and 3) the
temperature $T_g$ at which the structural relaxation time (from the self part 
of the intermediate scattering function) appears to
diverge. (b) The average disk positions at a temperature $T <
T_{cb}$ (gray-shaded disks). White solid lines indicate contacts
between disks in the backbone. The arrows represent the displacement of
the disks relative to their positions at $T=0$, where the length of each arrow is proportional to the logarithm of the displacement of the disk . (c) Same 
as in (b) except for the average disk positions at a temperature 
$T_{cb} < T < T_r$.  Gray-shaded disks without edges are rattlers,
circular outlines with dashed edges indicate 
the initial positions of rattler disks, and white-dotted lines 
show contacts that include 
rattlers. (d) Same as (c) except for the average disk positions 
at a temperature $T_r < T < T_g$.}
\label{dr_def}
\end{figure*}

\subsection{Measurement of the deviation of the specific heat and average positions}
\label{quantities}

In this section, we investigate the effects of form and
contact-breaking nonlinearities on two physical quantities: 1) the
deviation in the specific heat per particle at constant volume,
$\Delta {\overline C}_V$, from the value at $T=0$ and 2) the deviation
in the average disk positions $\Delta r$ from their positions at
$T=0$.  We will measure both quantities using constant energy MD
simulations with equal velocity-amplitude perturbations involving all
eigenmodes (i.e. $A_1 \omega^1=A_2 \omega^2=\ldots=A_k
\omega^k$). In general, nonlinearities will cause $\Delta
{\overline C}_V >0$ and $\Delta r >0$ for temperatures $T>0$.  Form
nonlinearities can occur for $T < T_{cb}$, while both form and
contact-breaking nonlinearities occur for $T > T_{cb}$.

\begin{figure}[h!]
\includegraphics[width=3in]{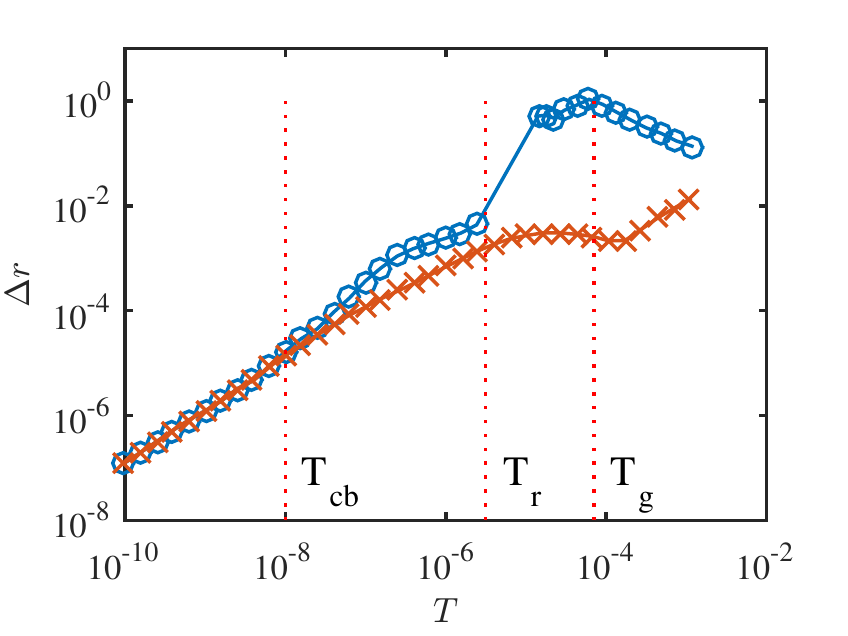}
\caption{The deviation $\Delta r$ in the disk positions from their
$T=0$ values as a function of temperature $T$ for a MS packing with
$N=32$, $U=10^{-5}$, and single- (circles) and double-sided (exes) 
Hertzian spring interactions. The three dotted vertical lines indicate
$T_{cb}$, $T_r$, and $T_g$ (from left to right).}
\label{dr_ratio_Hertz}
\end{figure}

\begin{figure*}
\includegraphics[width=17.5cm]{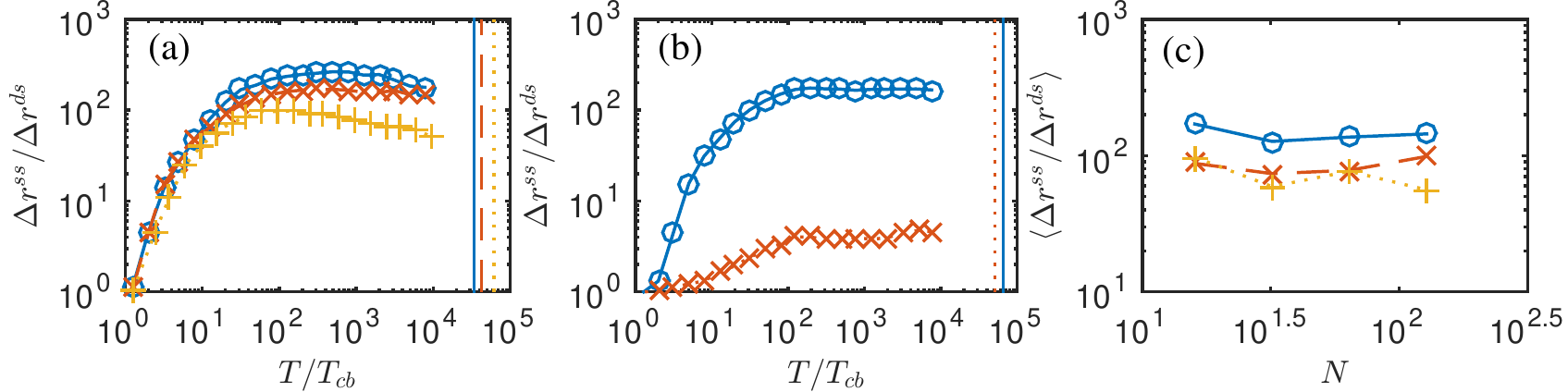}
\caption{(a) The ratio $\Delta r^{ss}/\Delta r^{ds}$ between the deviations
in positions for single- and double-sided linear spring interactions as a
function of temperature normalized by contact-breaking temperature
$T/T_{cb}$ for packings with $N=128$ and three values of $U$ ($10^{-5}$ 
(circles), $10^{-4}$ (exes), and
$10^{-3}$ (pluses)). Each curve
is averaged over $50$ packings in the temperature range $1 < T/T_{cb} < 10^4$.
The three vertical lines indicate $\langle {T}_r\rangle$ for these packings,
$U=10^{-5}$ (solid line), $10^{-4}$ (dashed line), and $10^{-3}$ (dotted 
line).  (b) The ratio of position deviations $\Delta
r^{ss}/\Delta r^{ds}$ from 
single- and double-sided linear (circles) and Hertzian (exes) 
spring interactions as
a function of $T/T_{cb}$ for MS packings with $N=32$ and 
$U=10^{-5}$.  Each curve
is averaged over $50$ packings in the temperature range $1 <
T/T_{cb} < 10^4$. The two vertical lines indicate $\langle
{T}_r\rangle$ for MS packings with purely repulsive linear (solid line) and
Hertzian spring interactions (dotted line). (c) $\langle \Delta r^{ss}/\Delta r^{ds} \rangle$ averaged over the 
temperature range $1 < T/T_{cb} < 10^4$ for linear spring interactions 
as a function of
system size $N$ for $U=10^{-5}$ (circles), $10^{-4}$ (exes), and
$10^{-3}$ (pluses). 
}
\label{dr_ratio}
\end{figure*}

We show $\Delta {\overline C}_V/{\overline C}_V^0$ (defined in
Eq.~\ref{dCv2}) as a function of temperature $T/T_{cb}$ (normalized by
the temperature $T_{cb}$ required to break a single contact) for
several system sizes for purely repulsive linear springs ($\alpha = 2$
in Eq.~\ref{Ep}) in Fig.~\ref{dCv} (a).  For purely repulsive linear
springs, the deviation $\Delta {\overline C}_V/{\overline C}_V^0$ is
set by the noise floor for $T < T_{cb}$, and thus deviations in the
specific heat per particle from form nonlinearities for $T < T_{cb}$
are below the noise floor. In Fig.~\ref{dCv} (b), we compare $\Delta
{\overline C}_V/{\overline C}_V^0$ for purely repulsive linear and
Hertzian springs ($\alpha = 5/2$ in Eq.~\ref{Ep}) as a function of
$T/T_{cb}$. As expected, the form nonlinearities are larger for
Hertzian interactions. In particular, the deviation in $\Delta
{\overline C}_V/{\overline C}_V^0$ is above the noise floor for
$T<T_{cb}$.

For purely repulsive linear springs, $\Delta {\overline
  C}_V/{\overline C}_V^0$ increases strongly above the noise floor for
temperatures near $T_{cb}$. $\Delta
{\overline C}_V/{\overline C}_V^0$ for purely repulsive Hertzian
springs also increases rapidly, but the onset of the rapid increase is
not as sharp and occurs for $T<T_{cb}$. However, the rate of increase
of $\Delta {\overline C}_V/C_V^0$ slows for increasing system sizes.
In the inset to Fig.~\ref{dCv} (a), we plot $\Delta {\overline
  C}_V/{\overline C}_V^0$ for $10$ values of $T/T_{cb}$ in the range
from $1$ to $10^2$ as a function of system size for purely repulsive
linear springs. We find that the deviation scales as $\Delta
{\overline C}_V/{\overline C}_V^0 \sim N^{-1}$ for a wide range of
$T/T_{cb}$, which implies that the effect of both form and contact-breaking 
nonlinearities on the specific
heat vanishes in the large-system limit in this temperature range.

We also study the change in the average disk positions $\Delta r$
(defined in Eq.~\ref{dr}) as a function of temperature using constant
energy MD simulations with equal velocity-amplitude initial
perturbations involving all eigenmodes.  We consider both purely
repulsive (single-sided) and double-sided linear and nonlinear spring
interactions. In Fig.~\ref{dr_def} (a) and~\ref{dr_ratio_Hertz},
we show $\Delta r^{ds}(T)$ (double-sided) and $\Delta r^{ss}(T)$
(single-sided) for disk packings with $N=64$, $U=10^{-5}$, and linear
and Hertzian spring interactions.  For double-sided linear and
Hertzian spring interactions, with no contact breaking, $\Delta r^{ds}
\sim T$ over a wide range of $T$.

This scaling behavior for $\Delta r^{ds}(T)$ stems from form
nonlinearities in the total potential energy ${\cal U}$, which when expanded
gives:
\begin{eqnarray}
\label{V_expand}
& & {\cal U} = {\cal U}^0 - \sum_i F^0_i\Delta R_i + \\
& & \frac{1}{2!} \sum_{i,j} D^0_{ij}\Delta R_i \Delta R_j+\frac{1}{3! }\sum_{i,j,k} G^0_{ijk}\Delta R_i \Delta R_j \Delta R_k+\ldots, \nonumber
\end{eqnarray}
where $\Delta {\vec R}=\vec{R}-\vec{R}^0$, $F^0_i=- \partial
V/\partial R_i |_{\Delta {\vec R}=0}$, $D^0_{ij}=\partial^2 V/
(\partial R_i \partial R_j)|_{\Delta {\vec R}=0}$, and
$G^0_{ijk}=\partial^3 V /(\partial R_i \partial R_j \partial
R_k)|_{\Delta {\vec R}=0}$. For $T < T_{cb}$, when the contact network does not change,
the third-order term in the expansion of ${\cal U}$ gives rise to the
scaling $\Delta r = C T$, where $C$ is set by $G^0$. (See
Appendix~\ref{dr_1d} for the calculation of $\Delta r$ for a potential 
with cubic terms in 1D.)  Rattler disks are excluded from the
measurement of $\Delta r$ because collisions between backbone and
rattler disks will introduce additional nonlinearities.  ($\Delta
r^{ss}$ for an MS packing with rattlers is shown in
Appendix~\ref{dr_flt}.)

As expected, for $T < T_{cb}$, $\Delta r^{ss} = \Delta r^{ds} \sim T$,
before contact breaking occurs for both linear and Hertzian spring
interactions. The disk displacements in this regime are small and
randomly oriented (Fig.~\ref{dr_def} (b)). For purely repulsive linear
spring interactions in the temperature regime $T > T_{cb}$, ${\Delta
  r}^{ss}$ begins to grow rapidly, reaching values that are several
orders of magnitude above $\Delta r^{ds}$. In the temperature regime
$T_{cb} < T < T_r$, some collective motion occurs and disks can
disconnect from the force-bearing backbone and become rattlers
(Fig.~\ref{dr_def} (c)). At $T=T_r$, $\Delta r^{ss}$ jumps
discontinuously when the system switches to the basin of a new MS
packing.  (See Appendix~\ref{rearrange} for a discussion of the method
that we used to measure $T_r$.) In the temperature regime $T_{r} < T <
T_{g}$, strong collective motion can occur and all of the disks can
disconnect from the force-bearing backbone when rattler disks are
identified recursively (Fig.~\ref{dr_def} (d)). Similar behavior
occurs for the deviations in the average positions for purely
repulsive Hertzian spring interactions (Fig.~\ref{dr_ratio_Hertz}),
i.e. $\Delta r^{ss}$ increases above $\Delta r^{ds}$ in the
temperature regime $T_{cb} < T < T_{r}$, but the increase is more
modest than that for repulsive linear springs.  
Comparing the disk positions at temperatures $T>T_g$ and zero is not
meaningful.

In Fig.~\ref{dr_ratio} (a), we plot the displacement ratio $\Delta
r^{ss}/\Delta r^{ds}$ for single- and double-sided linear spring
interactions as a function of $T/T_{cb}$ below $T_r$ for three values
of $U$ ($10^{-5}$ (circles), $10^{-4}$ (exes), and $10^{-3}$ (pluses))
and $N=128$. We find that the ratio begins growing for $T > T_{cb}$
reaching an approximate plateau value $\approx 100$ that increases
weakly with decreasing $U$.  Thus, contact-breaking nonlinearities are
much larger than form nonlinearities in the temperature range $T_{cb}
< T < T_r$ for linear spring interactions.  In Fig.~\ref{dr_ratio}
(b), we compare the ratio $\Delta r^{ss}/\Delta r^{ds}$ for linear and
Hertzian springs for $T_{cb} < T < T_r$. The contact breaking
nonlinearities have a much stronger effect on $\Delta r$ for linear
compared to Hertzian spring interactions.  This result likely stems
from the fact that form nonlinearities are much weaker for linear
spring interactions compared to Hertzian spring interactions. In
Fig.~\ref{dr_ratio} (c), we plot $\langle \Delta r^{ss}/\Delta r^{ds}
\rangle$ averaged over the temperature range $T_{cb} < T < T_r$ for
linear spring interactions as a function of system size $N$ for each $U$.
We find that $\langle \Delta r^{ss}/\Delta r^{ds} \rangle$ shows no
sign of decreasing with system size.  Thus, contact-breaking nonlinearities 
are dominant for MS packings with purely repulsive linear spring interactions 
in the temperature regime $T_{cb} < T < T_r$.

\section{Conclusions and Future Directions}
\label{conclusion}

In this article, we studied the effects of thermal fluctuations on MS
packings composed of bidisperse, frictionless disks generated at
different values of the potential energy per particle $U$ or excess
number of contacts $m/N$ in two spatial dimensions. We consider disks
that interact via single- and double-sided linear and nonlinear spring
interactions to disentangle the effects of form and contact-breaking
nonlinearities. To identify the temperature range where
contact-breaking nonlinearities occur, we first focused on calculating
the minimum temperature required to break a single contact in $T=0$ MS
packings for both single- and multi-mode perturbations.  Before
contact breaking and for weak form nonlinearities (e.g. purely
repulsive linear springs), the minimum temperature required to break a
single contact can be calculated exactly using the eigenmodes of the
dynamical matrix at $T=0$. Above the contact breaking temperature or
for interactions that possess strong form nonlinearities, the
eigenvalues and eigenmodes change significantly from those at $T=0$,
and thus the $T=0$ eigenvalues and eigenmodes cannot be used to
calculate the contact breaking temperature accurately.

For single eigenmode perturbations, we find that the minimum
temperature (over all single-mode excitations) required to break the
first contact, $T_1(m,m-1) \sim U/N^{\alpha}$, where $\alpha \approx
2.6$.  This strong system-size dependence emphasizes that weak
overlaps between disks in MS packings near jamming onset can break at any
finite temperature in the large-system limit. We also showed that the
form of the initial perturbation affects the minimum temperature
required to break a single contact.  The temperature required to break
a single contact is minimal for equal velocity-amplitude
perturbations involving all eigenmodes of the $T=0$ dynamical matrix
and scales as $T_n(m,m-1) \sim U/N^{\beta}$, where $\beta \sim
2.9$. $T_n(m,m-1)$ can be estimated by identifying the smallest pair
of overlapping disks $i$ and $j$ at a given $U$, shifting them so
that their separation satisfies $r_{ij} = \sigma_{ij}$, and then
minimizing the potential energy with $i$ and $j$ held fixed,
allowing the other disks to move.  The difference in the potential
energy per particle before ($U$) and after ($U'$) minimization $U'-U
\sim T_n(m,m-1)$ determines the minimum temperature required for
breaking a single contact for equal-velocity amplitude perturbations
involving all eigenmodes.

To study multiple contact breaking, we employed constant energy MD
simulations for initial packings at $U$ (and excess number of contacts
$m$) over a range of temperatures $T < T_r$.  We measure the fraction
of time during the simulations at a given temperature $T$ and system
size $N$ that the system possesses $N_{bc} = N^0_c + m - N_c$ broken
contacts.  We identify a characteristic temperature $T^*(N_{bc})$ at
which a finite fraction $f$ of the time (i.e.  $f=0.1$) the system
possesses a given number of broken contacts $N_{bc}$. By studying a
range of $U$ and $N$, we obtain the following power-law scaling
relation: $T^* \sim N_{bc}^{\gamma} N^{\delta} U^{\zeta}$, where
$\gamma \approx 2.2 \pm 0.3$, $\delta = -2.2 \pm 0.2$, and $\zeta
\approx 1.0 \pm 0.1$.  The scaling relation involving integer
exponents, $T^*/U \sim (N_{bc}/N)^2$, is within
error of the numerical data.  These results support prior 
studies that find that the temperature required to break 
an extensive number of contacts scales quadratically with 
the number of contact changes per particle. 

We also investigated the effects of form and contact-breaking
nonlinearities on the specific heat (at constant volume) and the
average disk positions as a function of temperature. We employed both
single- and double-sided linear and nonlinear spring interactions,
which allowed us to compare the strength of the form and
contact-breaking nonlinearities. For the specific heat per particle,
we find that the deviation $\Delta {\overline C}_V$ from the
zero-temperature value, ${\overline C}_V^0$, is below the noise
threshold for $T<T_{cb}$ for purely repulsive linear spring
interactions, and begins to increase rapidly for $T > T_{cb}$.  For
Hertzian interactions, the form nonlinearities give rise to measurable
deviations ${\overline C}_V^0/C_V^0$ for $T < T_{cb}$, and the strong
increase in ${\overline C}_V^0/C_V^0$ with increasing temperature
occurs over a larger range. However, we find that $\Delta {\overline
  C}_V/{\overline C}_V^0 \sim N^{-1}$ decreases with increasing system
size (for purely repulsive spring interactions) in the temperature
range $T_{cb} < T < T_r$.  Thus, we expect that form and contact-breaking
nonlinearities do not have strong effects on the specific heat for
$T<T_r$.

We also characterized the change in the average disk positions $\Delta
r$ from their $T=0$ values arising from form and contact-breaking
nonlinearities as a function of temperature.  $\Delta r$ is more
sensitive to form and contact-breaking nonlinearities than $\Delta
{\overline C}_V$. We first showed that $\Delta r^{ds} \sim T$ for
double-sided linear and Hertzian spring interactions over the full
range of temperature $0 < T < T_r$ due to form nonlinearities.  The
linear scaling with temperature arises from third-order terms in the
expansion of the total potential energy in terms of the disk
positions. As expected, $\Delta r^{ss} = \Delta r^{ds} \sim T$ for $T <
T_{cb}$ since there is no contact breaking.  Near $T= T_{cb}$, $\Delta
r^{ss}$ begins increasing rapidly above $\Delta r^{ds}$ for linear
springs due to contact-breaking nonlinearities.  We show that the
ratio $\Delta r^{ss}/\Delta r^{ds}$ can increase by a factor of $100$
for $T_{cb} < T < T_r$. In contrast, $\Delta r^{ss}/\Delta r^{ds} <
10$ for Hertzian interactions, presumably because the form
nonlinearities are much stronger. We show that $\Delta r^{ss}/\Delta
r^{ds}$ for linear springs displays very weak system size dependence.
This result emphasizes that contact-breaking nonlinearities are much
stronger than form nonlinearities for linear spring interactions in
this low-temperature regime.

Topics of future studies will include rattler disks, system
rearrangements, and nonlinearities induced by non-spherical particle
shapes.  In most of the current work, we excluded rattler disks 
by removing them from the MS packing before adding thermal
fluctuations.  As shown in Appendix~\ref{dr_flt}, additional
nonlinearities (e.g. collisions between disks in the $T=0$
force-bearing backbone and rattlers at $T>0$) are present when
rattlers are included in the system.  Second, in the current study, we
focused on the low-temperature regime $T < T_r$, below which the
system remains in the basin of the original $T=0$ MS packing.  In
future studies, we will characterize changes in key physical
quantities (such as the shear modulus) as the system moves among a
series of related basins for $T <T_g$, where the system is prevented
from undergoing complete structural relaxation~\cite{gardner}. The
current work was important in this context, since we characterized the
magnitude of changes in the disk positions that arise from
nonlinearities before rearrangements.

At low temperatures $T < T_{cb}$ and for systems with
weak nonlinearities, the eigenvalues and associated eigenmodes from
the dynamical matrix at $T=0$ agree with those from $S={\cal V} {\cal C}^{-1}$,
where ${\cal V}_{ij} = \langle v_i v_j \rangle$ is the time-averaged velocity 
correlation matrix and
\begin{equation}
\label{pos_cor}
{\cal C}_{ij} = \langle (R_i - R_i^0)(R_j - R_j^0)\rangle  
\end{equation}
is the time-averaged position correlation matrix~\cite{brito,henkes,bertrand2}. An important future
direction is to characterize how the eigenmodes of $S$ change as a
function of increasing temperature, e.g. do the modes become more or less
localized at a given frequency?

Another interesting research direction is to characterize the
nonlinearities that arise at finite temperature in MS packings
composed of non-spherical particles such as ellipsoids,
sphero-cylinders, or other elongated particles.  Several studies have
shown that packings of ellipsoids possess quartic modes near jamming
onset~\cite{schreck3,donev,zeravcic}, i.e. directions along which the
potential energy increases as the fourth power of the amplitude in
that direction.  These results point out that MS packings of
non-spherical particles possess form, contact-breaking, and {\it shape}
nonlinearities at finite temperature.  Determining the relative
strength of these nonlinearities and how they affect the structural
and mechanical properties of MS packings at finite temperature is an
important, open question.

\begin{acknowledgments}
The authors acknowledge financial support NSF grant nos.  CMMI-1462439
(C.O. and Q.W.), CMMI-1463455 (M.S.), and CBET-1605178 (C.O.  and
Q.W.).  This work was also supported by the High Performance Computing
facilities operated by, and the staff of, the Yale Center for Research
Computing.
\end{acknowledgments}

\appendix
\section{Calculation of minimum temperature required to break a single 
contact for equal velocity-amplitude perturbations} 
\label{Tstar}

In this Appendix, we provide additional details concerning the
calculation of the minimum temperature required to break a single
contact for perturbations involving multiple $T=0$ eigenmodes with
equal velocity-amplitude excitations. (See Sec.~\ref{Tcb}.)  In
Eq.~\ref{Tij}, we derived the expression for the minimum temperature
required to break a single contact (for $T < T_{cb}$ and systems with
weak nonlinearities) by setting $r^2_{ij} =\sigma^2_{ij}$ and using
Eq.~\ref{rt_n} for the time-dependent disk positions. Here, we will
justify why the the maximum of $r_{ij}^2$ is obtained when
$|\sin(\omega^1 t)|=|\sin(\omega^2 t)|=\ldots=|\sin(\omega^n t)|=1$, 
where $n$ is the number of eigenmodes in the initial perturbation.
The pair separations satisfy $r_{ij}^2 = x_{ij}^2+y_{ij}^2$, where
\begin{equation}
\label{xij}
x_{ij} = \Delta_x^0+\sum_{p=1}^n \Delta_x^p \sin(\omega^p t)
\end{equation}
\begin{equation}
\label{yij}
y_{ij} = \Delta_y^0+\sum_{p=1}^n \Delta_y^p \sin(\omega^p t),
\end{equation}
the parameters $\Delta_x^0$, $\Delta_x^1$,\ldots,$\Delta_x^n$, and
$\Delta_y^0$, $\Delta_y^1$,\ldots,$\Delta_y^n$ are constants
determined by the initial perturbation and positions of disks $i$ and
$j$.  We define $I^m_{ij} = (x^m_{ij})^2+(y^m_{ij})^2$, where
$x^m_{ij}= \Delta^0_x+\sum_{p=1}^m \Delta_x^p \sin(\omega^p t)$, and
$y^m_{ij}= \Delta_y^0+\sum_{p=1}^m \Delta^p_y \sin(\omega^p t)$. When
$m=0$, $I_m = (\Delta_x^0)^2+ (\Delta_y^0)^2$ and when $m=n$,
$I_m={r}_{ij}^2$. Suppose that when $m=q$, $I^q_{ij} =
(x^q_{ij})^2+(y^q_{ij})^2$ is maximal.  For $m=q+1$,
\begin{eqnarray}
\label{Iq}
& & I^{q+1}_{ij} = \\
& & (x_{ij}^q + \Delta_{x}^{q+1} \sin(\omega^{q+1} t))^2+(
y^{q}_{ij}+\Delta_y^{q+1} \sin(\omega^{q+1} t))^2.\nonumber
\end{eqnarray}

\begin{figure}[h!]
\includegraphics[width=3in]{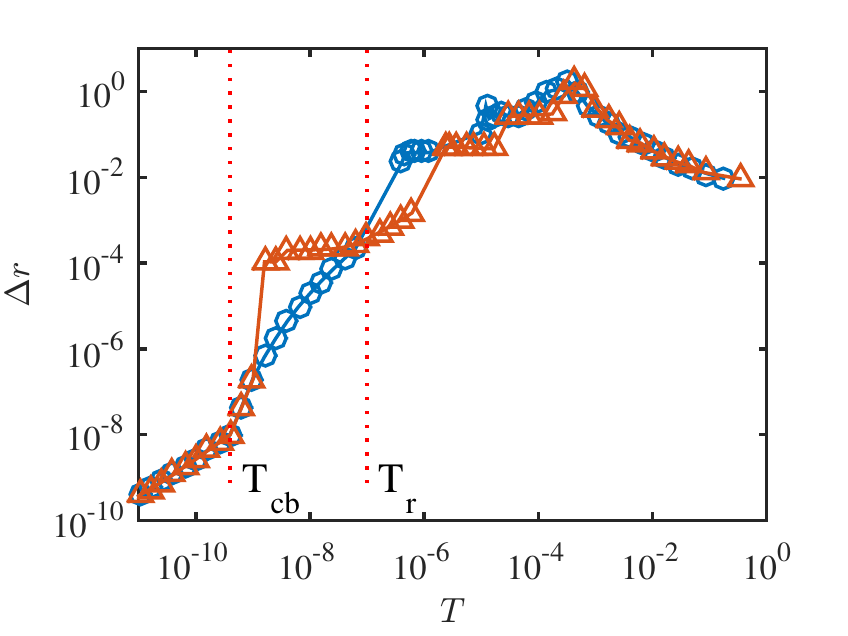}
\caption{$\Delta r$ versus temperature $T$ for an initial   
MS packing with purely repulsive linear spring interactions, $N=64$, 
and $U=10^{-5}$ with the two rattlers kept in the system (triangles)
and the two rattlers removed (circles).}
\label{dr_fltNoflt}
\end{figure}

The maximum of $I^{q+1}_{ij}$ is obtained when $dI^{q+1}_{ij}/dt=0$, which 
is satisfied when $\cos(\omega^{q+1}t)=0$ and
$|\sin(\omega^{q+1} t)|=1$. When the proof by induction is repeated, 
the maximum $r_{ij}^2$ is obtained if and only if
$|\sin(\omega^1 t)|=|\sin(\omega^2 t)|=\ldots=|\sin(\omega^n
t)|=1$. We then study all possible combinations of $\pm 1$ for 
each of the sine terms and and disk pairs $i$ and $j$ and choose those that 
give the smallest perturbation temperature. 

\section{Measurement of $\Delta r$ in MS packings with rattlers}
\label{dr_flt}

In Fig.~\ref{dr_def}, we showed results for $\Delta r$ (the deviation
of the average positions of the disks from their $T=0$ values) using
constant energy MD simulations as a function of temperature for MS
packings with rattlers removed from the system before the
perturbations were applied. In this Appendix, we show that rattlers
can have a strong effect on $\Delta r$ by introducing new
nonlinearities into the system.  In Fig.~\ref{dr_fltNoflt}, we compare
$\Delta r(T)$ for an MS disk packing with the same force-bearing
backbone at $T=0$ (with purely repulsive linear spring interactions)
with and without rattlers removed.  (Note that the rattlers do not
directly receive thermal excitations.)  For sufficiently low
temperatures when the rattlers are not excited by fluctuations in the
force-bearing backbone, $\Delta r(T)$ is the same for both systems
with and without rattlers. For the MS packing studied in
Fig.~\ref{dr_fltNoflt}, the force-bearing backbone comes into contact
with the rattlers at a temperature slightly above $T_{cb}$ (defined
using the force-bearing backbone at $T=0$) and $\Delta r$ jumps
discontinuously for the system with rattlers. (Note that the jump in
$\Delta r$ can occur over a range of temperatures depending on the
placement of the rattlers.) Above this temperature, the evolution of
$\Delta r$ is different for the systems with and without rattlers,
until the system without rattlers switches to the basin of a new MS
packing.  Since this article focused on quantifying form and
contact-breaking nonlinearities in the temperature regime $T_{cb} < T
< T_r$, we mainly performed MD simulations of MS packings with
rattlers removed.

\section{Measurement of the rearrangement and glass transition 
temperatures, $T_r$ and $T_g$}
\label{rearrange}

Our constant energy MD simulations mainly focused on the temperature
regime $T_{cb} < T < T_r$, where $T_{cb}$ is the temperature at which
the first contact breaks during the simulations and $T_r$ is the
temperature below which the system remains in the basin of the $T=0$
MS packing. To calculate $T_r$, we first simulate a long trajectory at
a temperature $T$ for a given initial perturbation and total time
$t_{tot}$.  For each time step of the simulation, we use the current
configuration as the initial condition for finding the nearest MS
packing (at a given $U$) using the packing-generation protocol described in
Sec.~\ref{methods}.  We then calculate the fraction $f_i$ of time that 
the system spends in the basin of MS packing $i$.  The MS packings 
are distinguished using the eigenvalues of the dynamical matrix. 

\begin{figure}[h!]
\includegraphics[width=3in]{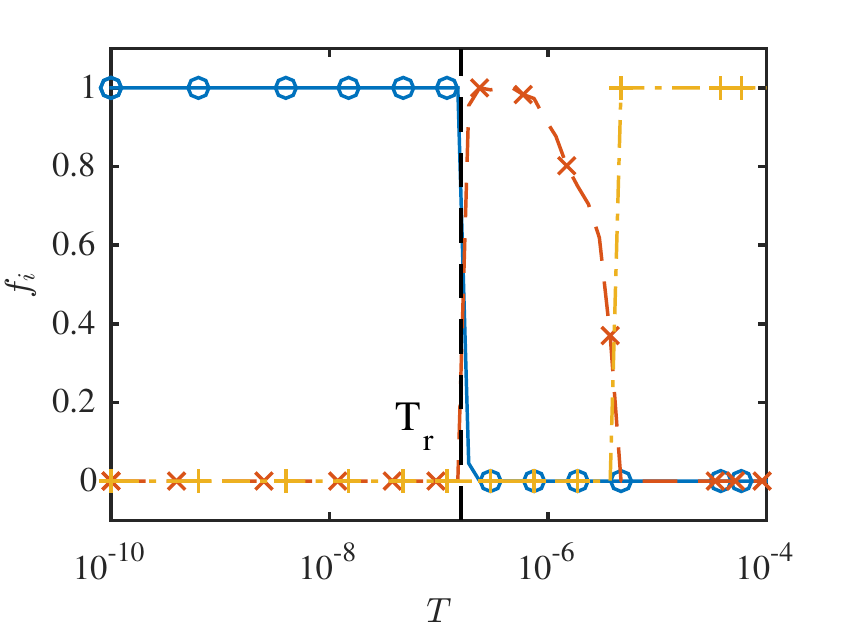}
\caption{The fraction $f_i$ of time that three particular MS packings occur 
in the temperature range $10^{-10} < T < 10^{-4}$ for systems with $N=64$ and $U=10^{-5}$. The
dashed vertical line indicates the rearrangement temperature $T_r$ for the 
$T=0$ MS packing. At the lowest temperatures, the system only 
populates the $T=0$ MS packing (circles). At intermediate temperatures 
a different MS packing (exes) 
becomes most frequent. At the highest temperatures, the system 
spends all of the time in a third MS packing 
(pluses).}
\label{Tr}
\end{figure}

In Fig.~\ref{Tr}, we plot $f_i$ as a function of temperature $T$ after
perturbing a given $T=0$ MS packing with equal velocity-amplitude
excitations involving all eigenmodes at each $T$. We find that three
particular MS packings occur most frequently over this range of $T$ and for this
initial condition.  At the lowest $T$, only the $T=0$ MS packing
(circles) occurs.  At $T_r$, the fraction of time that the system
spends in the $T=0$ MS packing tends to zero, and the fraction of time
that the system spends in a new MS packing (exes) increases to one.
At $T \approx 10^{-6}$, the system begins spending time in several MS
packings, and at $T \approx 10^{-5}$, the system spends all of its
time in a third MS packing (pluses).  In most cases, the behavior of
$f_i(T)$ mimics that shown in Fig.~\ref{Tr} for the first
rearrangement, i.e. there is a rapid drop in occupancy of the $T=0$ MS
packing and a rapid increase in the occupancy of another MS packing at
a well-defined temperature.  Thus, $T_r$ can be measured accurately
for each $T=0$ MS packing.  We also find strong agreement when we
measure $T_r$ using $f_i$ and when we define $T_r$ as the temperature
at which the first discontinuous jump in $\Delta r$ occurs for systems 
where rattlers have been removed. (See
Fig.~\ref{dr_def} (a).)

\begin{figure}[h!]
\includegraphics[width=3in]{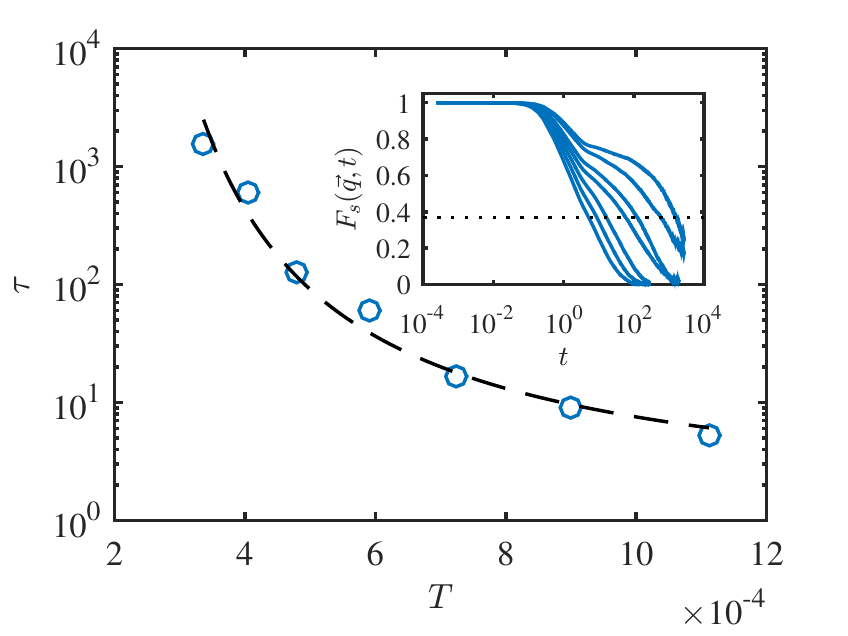}
\caption{Structural relaxation time $\tau$ (from the decay of the 
self-part of the intermediate scattering function) as a function of
temperature $T$ for a system with $N=64$ and $U=10^{-5}$. The dashed
line gives $\tau(T)=C \exp[A T_g/(T-T_g)]$, where $C=1.1$,
$A=15$, and $T_g=1.3 \times 10^{-4}$. The inset shows the self-part of the 
intermediate 
scattering function at $q\sigma_S = 2\pi$, $F_s(q,t)$, for several 
temperatures from $T=10^{-4}$ to $10^{-3}$ from top to
bottom. The horizontal line
indicates $F_s(q,\tau)=e^{-1}$.}
\label{tau_T}
\end{figure}

To emphasize that our measurements focus on the extremely
low-temperature regime, we also calculated the structural relaxation
time from the self-part of the intermediate scattering function (ISF) versus 
temperature~\cite{kob}:
\begin{equation}
\label{ISF}
F_s(\vec{q}, t) = \frac{1}{N} \sum_{j=1}^N \langle \exp(-i\vec{q}\cdot[\vec{r}_j(t)-\vec{r}_j(0)])\rangle, 
\end{equation} 
where $\vec{q}$ is the wave number and $\langle \cdot \rangle$
indicates an average over time origins and directions of the
wavevector. Near the glass transition temperature, the ISF develops a
plateau, whose length increases dramatically with decreasing $T$. At
the longest timescales and for $T > T_g$, the ISF decays as a
stretched exponential with stretching exponents that depend on $q$ and 
$T$~\cite{stretching}. (See the inset to Fig.~\ref{tau_T}.) We define a structural relaxation
time $\tau$ as $F_s(q,\tau) = e^{-1}$ for $q\sigma_S =2\pi$.

For fragile glasses, the structural relaxation time obeys 
super-Arrhenius scaling with temperature~\cite{angell}.  As a rough estimate 
of the glass transition temperature $T_g$, we use the Vogel-Fulcher-Tammann 
form~\cite{goldstein} for $\tau(T)$: 
\begin{equation}
\label{tau}
\tau \sim \exp[A T_g/(T-T_g)],
\end{equation}
where $A$ is a constant and $T_g$ is glass transition temperature at 
which the structural relaxation time appears to diverge. In Fig.~\ref{tau_T}, 
we show that for $N=64$ and $U=10^{-5}$, $T_g \approx 10^{-4}$, which 
is several orders of magnitude larger than $T_{cb}$ and $T_r$ for this system.

\begin{figure}[h!]
\includegraphics[width=3in]{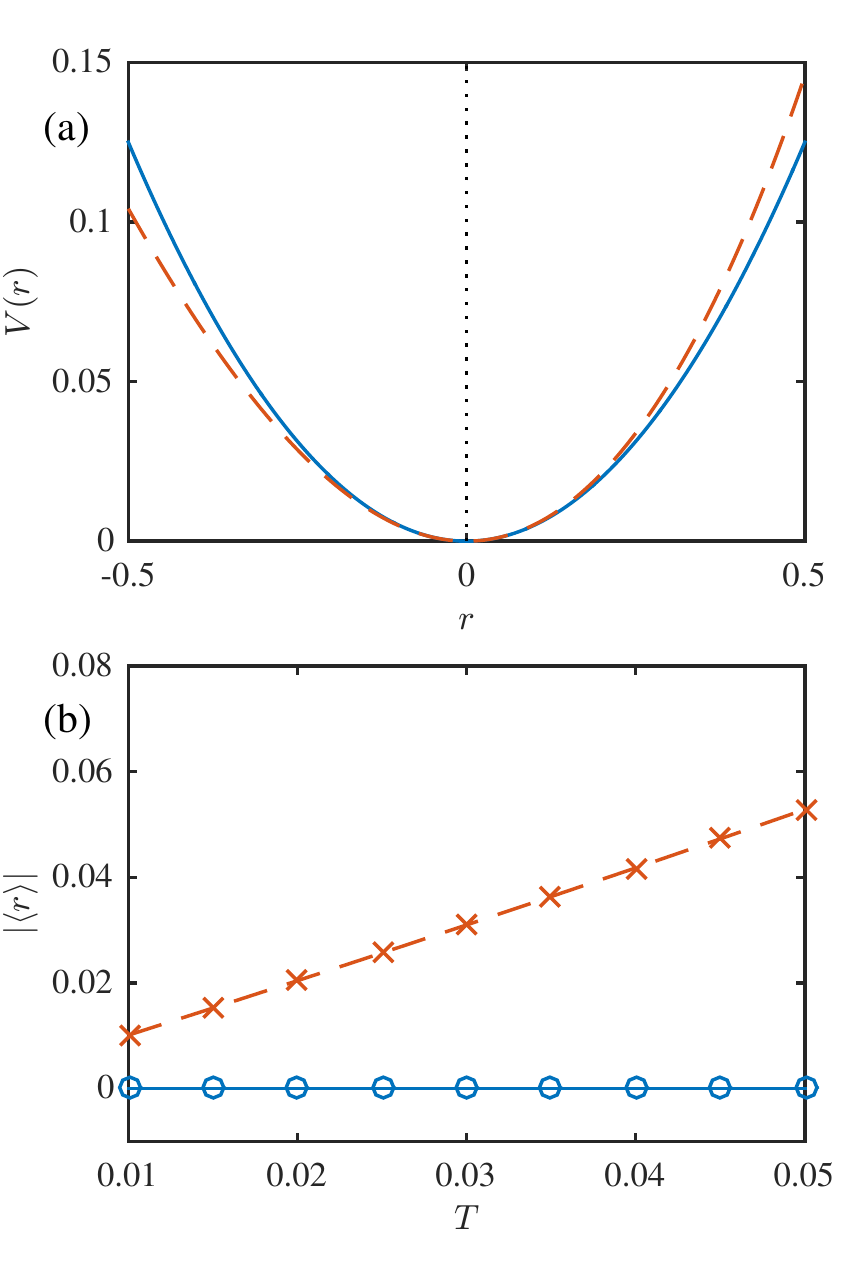}
\caption{(a) The potential energy $U(r)$ as a function of position $r$ 
for a quadratic form, $U^q(r)=Ar^2/2$ (solid line),
and a cubic form, $V^c(r)=Ar^2/2+Br^3/6$ (dashed line). The vertical
dotted line indicates $r=0$. Note that the cubic potential is asymmetric 
about $r=0$. (b) The absolute value of the average position $|\langle r\rangle|$
versus temperature $T$ for the quadratic (circles) and cubic (exes)
potentials.}
\label{1D_dr}
\end{figure}

\section{The temperature dependence of $\Delta r$ in model 1D systems}
\label{dr_1d}

To better understand the temperature scaling of the average position 
deviation, $\Delta r \sim T$, for
MS packings at non-zero temperatures, we studied a model system
consisting of a particle in a one-dimensional (1D) potential well. We
considered two forms for the potential: a quadratic potential,
$U^q(r)= Ar^2/2$, and a cubic potential, $U^c(r) = Ar^2/2 + Br^3/6$ as
shown in Fig.~\ref{1D_dr} (a).
   
The average position $\langle r\rangle$ as a function 
of temperature can be calculated using
\begin{equation}
\label{r_mean_1d}
\langle r\rangle=\frac{\int_0^{\infty} r f(r) dr}{\int_0^{\infty} f(r) dr},
\end{equation}
where the position distribution function in 1D is
\begin{equation}
\label{fr_1d}
f(r)=\frac{1}{\sqrt{2(2T-U(r))}}.
\end{equation}

For the quadratic potential, the average particle position $\langle
r\rangle = 0$ for all $T$.  In contrast, for the cubic potential,
$|\langle r\rangle| = B T/A^2$ increases linearly with $T$ with a slope
that scales with the coefficient of the cubic term. A similar analysis
can be applied to MS packings of disks. Before contact breaking, the
system lies in a high-dimensional potential energy well. All of the
potentials that we studied (i.e. Eqs.~\ref{Ep} and~\ref{double} with
$\alpha = 2$ and $5/2$) possess ``form'' nonlinearities with nonzero
values for the third derivatives of the total potential energy with
respect to the disk positions (Eq.~\ref{V_expand}). Thus, similar to
the model 1D system, $\Delta r$ in MS packings before contact breaking is
proportional to the temperature $T$ with a slope that is determined by
the third-derivative of the potential energy with respect to the
particle coordinates in the direction of the initial perturbation.

\bibliography{contact_breaking.bib}

\end{document}